\documentclass[useAMS,usenatbib]{mn2e/mn2e}

\usepackage{graphicx}
\usepackage{txfonts}
\usepackage{xcolor}
\usepackage{aecompl}
\usepackage{supertabular,booktabs}
\usepackage{float}
\usepackage{amssymb}
\usepackage{hyperref}
\hypersetup{
         colorlinks = True, 
     linkcolor = {blue},
     linkbordercolor = {blue},
     anchorcolor=green,
     citecolor= {blue} }

\newcommand{\alpfe}{~[$\alpha$/Fe]~}
\newcommand{\msun}{M$_\odot$}

\newcommand{\parsec}{~\texttt{PARSEC}~}
\newcommand{\gaia}{\textit{Gaia}~}
\newcommand {\reac}[6] {$\rm\,{}^{#2}\kern-0.8pt{#1}\,({#3}\,,{#4})
  \,{}^{#6}\kern-0.8pt{#5}\,$}

\title[\parsec $\alpha$-enhanced isochrones: 47 Tuc and RGBB]
{New \parsec database of $\alpha$-enhanced stellar evolutionary tracks and isochrones I. 
Calibration with 47~Tuc (NGC104) and the improvement on RGB bump}

\author[X. Fu, et al.]
{
\parbox{\textwidth}{Xiaoting Fu$^{1}$,
Alessandro Bressan$^{1}$,
Paola Marigo$^{2}$,
L\'{e}o Girardi$^{3}$,
Josefina Montalb\'an$^{2}$,
Yang Chen$^{2}$,
Ambra Nanni$^{2}$}
\vspace{0.4cm}\\
\parbox{\textwidth}{
        $^1$ SISSA - International School for Advanced Studies, via Bonomea 265, 34136 Trieste, Italy \\
$^2$ Dipartimento di Fisica e Astronomia Galileo Galilei, Universit\`a di Padova, Vicolo dell'Osservatorio 3, I-35122 Padova, Italy \\
$^3$ INAF - Osservatorio Astronomico di Padova, Vicolo dell'Osservatorio 5, I-35122 Padova, Italy }}

\begin{document}

\pagerange{\pageref{firstpage}--\pageref{lastpage}} \pubyear{2016}

\maketitle

\label{firstpage}

\begin{abstract}
        Precise studies on the Galactic bulge,  globular cluster, Galactic halo
        and Galactic thick disk require stellar models with $\alpha$
        enhancement and various values of  helium content.  These models are
        also important for extra-Galactic population synthesis studies.  For
        this purpose we complement the existing \parsec models, which are based
        on the solar partition of heavy elements, with $\alpha$-enhanced
        partitions.  We collect detailed measurements on the metal mixture and
        helium abundance for the two populations of 47~Tuc (NGC~104) from the
        literature, and calculate  stellar tracks and isochrones with these
        $\alpha$-enhanced compositions.  By fitting the precise color-magnitude
        diagram with \textit{HST ACS/WFC} data, from low main sequence till
        horizontal branch, we calibrate some free parameters that are important
        for the evolution of low mass stars like the mixing at the bottom of
        the convective envelope.  This new calibration significantly improves
        the prediction of the RGB bump brightness.  Comparison with the
        observed RGB and HB luminosity functions also shows that the
        evolutionary lifetimes are correctly predicted.  As a further result of
        this calibration process, we derive the age, distance modulus,
        reddening, and the  red giant branch mass loss for 47~Tuc.  We apply
        the new calibration and $\alpha$-enhanced mixtures of the two 47~Tuc
        populations (\alpfe $\sim$0.4 and 0.2) to other metallicities.  The new
        models reproduce the RGB bump observations much better than previous
        models.  This new \parsec database, with the newly updated
        $\alpha$-enhanced stellar evolutionary tracks and isochrones, will also
        be part of the new stellar products for ~\gaia.

\end{abstract}


\section{Introduction}

\parsec (PAdova-TRieste Stellar Evolution Code) is widely used in the
astronomical community. It provides input for population synthesis models to
study resolved and unresolved star clusters and galaxies
\citep[e.g.][]{Perren2015, Gutkin2016, Chevallard2016}, and offers reliable
models for many other field of studies, such as to derive black hole mass when
observing gravitational waves \citep[e.g.][]{Spera2015, Belczynski2016}, to get
host star parameters for exoplanets \citep[][etc.]{Santos2013, Maldonado2015},
to explore the mysterious ``cosmological lithium problem'' \citep{pms}, to
derive the main parameters of star clusters \citep[for instance,][]{Donati2014,
Borissova2014, Roman2015, Goudfrooij2015} and Galactic structure
\citep[e.g.][]{kupper2015, Li2016, Balbinot2016, Ramya2016}, to study dust
formation \citep[e.g.][]{Nanni2013, Nanni2014}, to constrain dust extinction
\citep[e.g.][]{Schlafly2014,Schultheis2015, Bovy2016}, and to understand the
stars themselves \citep[e.g.][]{Kalari2014, Smiljanic2016, Gullikson2016,
Reddy2016, Casey2016}, etc.

There are now four versions of \parsec isochrones available online\footnote{CMD
input form: \url{http://stev.oapd.inaf.it/cgi-bin/cmd/}}. The very first
version \parsec \texttt{v1.0} \citep{Bressan2012} provides isochrones for
0.0005$\leq$Z$\leq$0.07 ($-$1.5$\leq$[M/H]$\leq$+0.6) with the mass range
0.1~\msun$\leq$M$<$12~\msun ~from pre-main sequence to the thermally pulsing
asymptotic giant branch (TP-AGB).   In \parsec \texttt{v1.1} \citep[based on
][]{Bressan2012} we expanded the metallicity range down to Z=0.0001
([M/H]=$-$2.2).  \parsec \texttt{v1.2S} included big improvements both on the
very low mass stars and massive stars: \citet{Chen2014} improve the surface
boundary conditions for stars with mass $M\lesssim$ 0.5~\msun~ in order to fit
the mass-radius relation of dwarf stars; \citet{Tang2014} introduce mass loss
for massive star M$\geq$14~\msun; \citet{Chen2015} improve the  mass-loss rate
when the luminosity approaches the Eddington luminosity and supplement the
model with new bolometric corrections till M=350~\msun.  In a later version
(\parsec \texttt{v1.2S + COLIBRI PR16}) we describe improved isochrones with
the addition of COLIBRI \citep{Marigo2013} evolutionary tracks of TP-AGB stars
\citep{Rosenfield2016, Marigo2017}.

All previous versions of \parsec models are calculated  assuming solar-scaled
metal mixtures, in which the initial partition of heavy elements keeps always
the same relative number density as that in the Sun.  It is now well
established that the solar-scaled metal mixture is not universally applicable
for all types of stars.  In fact, one of the most important group of elements,
the so called $\alpha$-elements group, is not always observed in solar
proportions.  Many studies have  confirmed the existence of an ``enhancement''
of $\alpha$-elements in the Milky Way halo \citep[e.g.][]{Zhao1990,Nissen1994,
McWilliam1995, Venn2004}, globular clusters  \citep[e.g.][]{Carney1996,
Sneden2004, Pritzl2005}, the Galactic Bulge \citep{Gonzalez2011, Johnson2014},
and  thick disk \citep[e.g.][]{Fulbright2002, Reddy2006, Ruchti2010}.  Stars in
the dwarf spheroidal Milky Way satellite galaxies show different
$\alpha$-abundance trends compared to the Galactic halo stars, possibly
indicating different star formation paths \citep{Kirby2011}.  The
$\alpha$-elements (O, Ne, Mg, Si, S, Ar, Ca, and Ti) are mainly produced by
core collapse (mostly Type II)  supernovae (SNe) on short timescale, while the
iron-peak elements (V, Cr, Mn, Fe, Co and Ni)  are mainly synthesized in Type
Ia SNe on longer timescales.  Therefore, the evolution profile of \alpfe
records the imprint of the star formation history of the system.  An
alternative explanation could be that the Initial Mass Function (IMF) of the
$\alpha$-enhanced stellar populations was much richer in massive stars than the
one from which our Sun was born \citep{Chiosi1998}.  However there is no clear
evidence in support of this alternative possibility.

In order to model star clusters, galaxies and Galactic components more
precisely, the previous Padova isochrone database offered a few sets of
$\alpha$-enhanced models for four relatively high metallicities
\citep{Salasnich2000}, other stellar evolution  groups also published
isochrones that allow  for $\alpha$ enhancement \citep[e.g.][]{VandenBerg2000,
VandenBerg2014, Pietrinferni2006, Valcarce2012}.  Now, with the thorough
revision and update input physics, we introduce $\alpha$-enhanced metal
mixtures in \parsec.

In this paper we first calibrate the new \parsec $\alpha$-enhanced stellar
evolutionary tracks and isochrones with the well-studied globular cluster
47 Tucanae (NGC 104), Then we apply the calibrated parameters to obtain
models for other metallicities.  Section \ref{sec:input} briefly describes
the input physics.  Section \ref{sec:47tuc} introduces the comparison with
47~Tuc data in details, including the isochrone fitting and luminosity
function, envelope overshooting calibration with red giant branch bump, and
mass loss in the red giant branch (RGB) from horizontal branch (HB)
morphology.  Section \ref{sec:com} compares the new \parsec models with
other stellar models and shows its improvement on RGB bump prediction.  A
summary of this paper and the discussion are in Section \ref{sec:conc}.

\section{Input physics}
\label{sec:input}

The main difference with respect to the previous versions of \parsec is the
adoption of new nuclear reaction rates, $\alpha$-enhanced opacities, and
various helium contents.

We update the nuclear reaction rates from JINA REACLIB database \citep{jina}
with their  April 6, 2015 new recommendation.  In addition, more reactions --
52 instead of the 47 described in \citet{Bressan2012} for the previous versions
of \parsec\ -- are taken into account.  They are all listed in
Table~\ref{tab:reac} together with the reference from which we take the
reaction energy $Q$ value.  In the updated reactions, more isotopic abundances
are considered, in total $N_{el}=29$: $^1$H, D, $^3$He, $^4$He,$^7$Li, $^8$Be,
$^8$B,  $^{12}$C, $^{13}$C, $^{14}$N, $^{15}$N, $^{16}$N, $^{17}$N, $^{17}$O,
$^{18}$O, $^{18}$F, $^{19}$F, $^{20}$Ne, $^{21}$Ne, $^{22}$Ne, $^{23}$Na,
$^{24}$Mg, $^{25}$Mg, $^{26}$Mg, $^{26}$Al$^m$, $^{26}$Al$^g$, $^{27}$Al,
$^{27}$Si, and $^{28}$Si.

    \begin{table}
            \centering
            \caption{Nuclear reaction rates adopted in this work and the reference from which we take their reaction energy $Q$.}
            \begin{tabular}{@{}ll@{}}
                    \hline
                    \multicolumn{1}{c}{Reaction} &
                    \multicolumn{1}{c}{Reference} \\
                    \hline
                    \reac{p}{}{p}{\beta^+\,\nu}{D}{} & \citet{bet}  \\
                    \reac{p}{}{D}{\gamma}{He}{3} &  \citet{de04}\\
                    \reac{He}{3}{^{3}He}{\gamma}{2\,p + ^{4}\kern-0.8pt{He}}{} &  \citet{nacre}\\
                    \reac{He}{4}{^{3}He}{\gamma}{Be}{7} &  \citet{cd08}\\
                    \reac{Be}{7}{e^-}{\gamma}{Li}{7} &  \citet{jina}\\
                    \reac{Li}{7}{p}{\gamma}{^{4}\kern-2.0pt{He} + ^{4}\kern-2.0pt{He}}{} &  \citet{de04} \\
                    \reac{Be}{7}{p}{\gamma}{B}{8} & \citet{nacre}  \\
                    \reac{C}{12}{p}{\gamma}{N}{13} & \citet{ls09}  \\
                    \reac{C}{13}{p}{\gamma}{N}{14} & \citet{nacre}  \\
                    \reac{N}{14}{p}{\gamma}{O}{15} & \citet{im05}  \\
                    \reac{N}{15}{p}{\gamma}{^4He + ^{12}\kern-2.0pt{C}}{} & \citet{nacre}  \\
                    \reac{N}{15}{p}{\gamma}{O}{16} & \citet{il10}  \\
                    \reac{O}{16}{p}{\gamma}{F}{17} & \citet{ia08}  \\
                    \reac{O}{17}{p}{\gamma}{\,^4He + ^{14}\kern-2.0pt{N}}{} & \citet{il10}  \\
                    \reac{O}{17}{p}{\gamma}{F}{18} & \citet{il10}  \\
                    \reac{O}{18}{p}{\gamma}{\,^4He + ^{15}\kern-2.0pt{N}}{} & \citet{il10}  \\
                    \reac{O}{18}{p}{\gamma}{F}{19} & \citet{il10}  \\
                    \reac{F}{19}{p}{\gamma}{\,^4He + ^{16}\kern-2.0pt{O}}{} & \citet{nacre}  \\
                    \reac{F}{19}{p}{\gamma}{Ne}{20} & \citet{nacre}  \\
                    \reac{He}{4}{2\,^{4}He}{\gamma}{C}{12} & \citet{fy05}  \\
                    \reac{C}{12}{^{4}He}{\gamma}{O}{16} & \citet{chw0}  \\
                    \reac{N}{14}{^{4}He}{\gamma}{F}{18} & \citet{il10}  \\
                    \reac{N}{15}{^{4}He}{\gamma}{F}{19} & \citet{il10}  \\
                    \reac{O}{16}{^{4}He}{\gamma}{Ne}{20} & \citet{co10}  \\
                    \reac{O}{18}{^{4}He}{\gamma}{Ne}{22} & \citet{il10}  \\
                    \reac{Ne}{20}{^{4}He}{\gamma}{Mg}{24} & \citet{il10}  \\
                    \reac{Ne}{22}{^{4}He}{\gamma}{Mg}{26} & \citet{il10}  \\
                    \reac{Mg}{24}{^{4}He}{\gamma}{Si}{28} & \citet{st08}  \\
                    \reac{C}{13}{^{4}He}{n}{O}{16} & \citet{hd08}  \\
                    \reac{O}{17}{^{4}He}{n}{Ne}{20} & \citet{nacre}  \\
                    \reac{O}{18}{^{4}He}{n}{Ne}{21} & \citet{nacre}  \\
                    \reac{Ne}{21}{^{4}He}{n}{Mg}{24} & \citet{nacre}  \\
                    \reac{Ne}{22}{^{4}He}{n}{Mg}{25} & \citet{il10}  \\
                    \reac{Mg}{25}{^{4}He}{n}{Si}{28} & \citet{nacre}  \\
                    \reac{Ne}{20}{p}{\gamma}{Na}{21} & \citet{il10}  \\
                    \reac{Ne}{21}{p}{\gamma}{Na}{22} & \citet{il10}  \\
                    \reac{Ne}{22}{p}{\gamma}{Na}{23} & \citet{il10}  \\
                    \reac{Na}{23}{p}{\gamma}{\,^4He + ^{20}\kern-2.0pt{Ne}}{} & \citet{il10}  \\
                    \reac{Na}{23}{p}{\gamma}{Mg}{24} & \citet{il10}  \\
                    \reac{Mg}{24}{p}{\gamma}{Al}{25} & \citet{il10}  \\
                    \reac{Mg}{25}{p}{\gamma}{Al^g}{26} & \citet{il10}  \\
                    \reac{Mg}{25}{p}{\gamma}{Al^m}{26} & \citet{il10}  \\
                    \reac{Mg}{26}{p}{\gamma}{Al}{27} & \citet{il10}  \\
                    \reac{Al^g}{26}{p}{\gamma}{Si}{27} & \citet{il10}  \\
                    \reac{Al}{27}{p}{\gamma}{\,^4He + ^{24}\kern-2.0pt{Mg}}{} & \citet{il10}  \\
                    \reac{Al}{27}{p}{\gamma}{Si}{28} & \citet{il10}  \\
                    \reac{Al}{26}{p}{\gamma}{Si}{27} & \citet{il10}  \\
                    \reac{Al}{26}{n}{p}{Mg}{26} & \citet{wc12}  \\
                    \reac{C}{12}{^{12}C}{n}{Mg}{23}& \citet{cf88}  \\
                    \reac{C}{12}{^{12}C}{p}{Na}{23}& \citet{cf88}  \\
                    \reac{C}{12}{^{12}C}{^{4}He}{Ne}{20}& \citet{cf88}  \\
                    \reac{Ne}{20}{\gamma}{^{4}He}{O}{16}& \citet{co10}  \\
                    \hline
                    \label{tab:reac}
            \end{tabular}
    \end{table}

The $\alpha$-enhanced opacities and equation of state (EOS) are  derived
for our best estimate of the  metal mixture of 47~Tuc, which is described
in Sec.~\ref{subsec:met}.  Details about the preparation of the opacity
tables are provided in \citet{Bressan2012}.  Suffice it to recall that
the Rosseland mean opacities come from two sources: from the Opacity
Project At Livermore \citep[OPAL,][and references
therein]{Iglesias1996}\footnote{\url{http://opalopacity.llnl.gov/}} team
at high temperatures ($4.0<\log(T/{\rm K})<8.7$), and from
\texttt{AESOPUS}
\citep{MarigoAringer2009}\footnote{\url{http://stev.oapd.inaf.it/aesopus}}
at low temperatures ($3.2<\log(T/{\rm K})<4.1$), with a smooth transition
being adopted in the $4.0<\log(T/{\rm K})<4.1$ interval.  Conductive
opacities are provided by \citet{Itoh2008} routines.  As for the EOS, we
choose the widely used  \texttt{FreeEOS} code (version 2.2.1 in the EOS4
configuration) \footnote{\url{http://freeeos.sourceforge.net/}} developed
by Alan W.\ Irwin for its computational efficiency.

\begin{figure}
              \centering
              \includegraphics[width=.48\textwidth,angle=0]{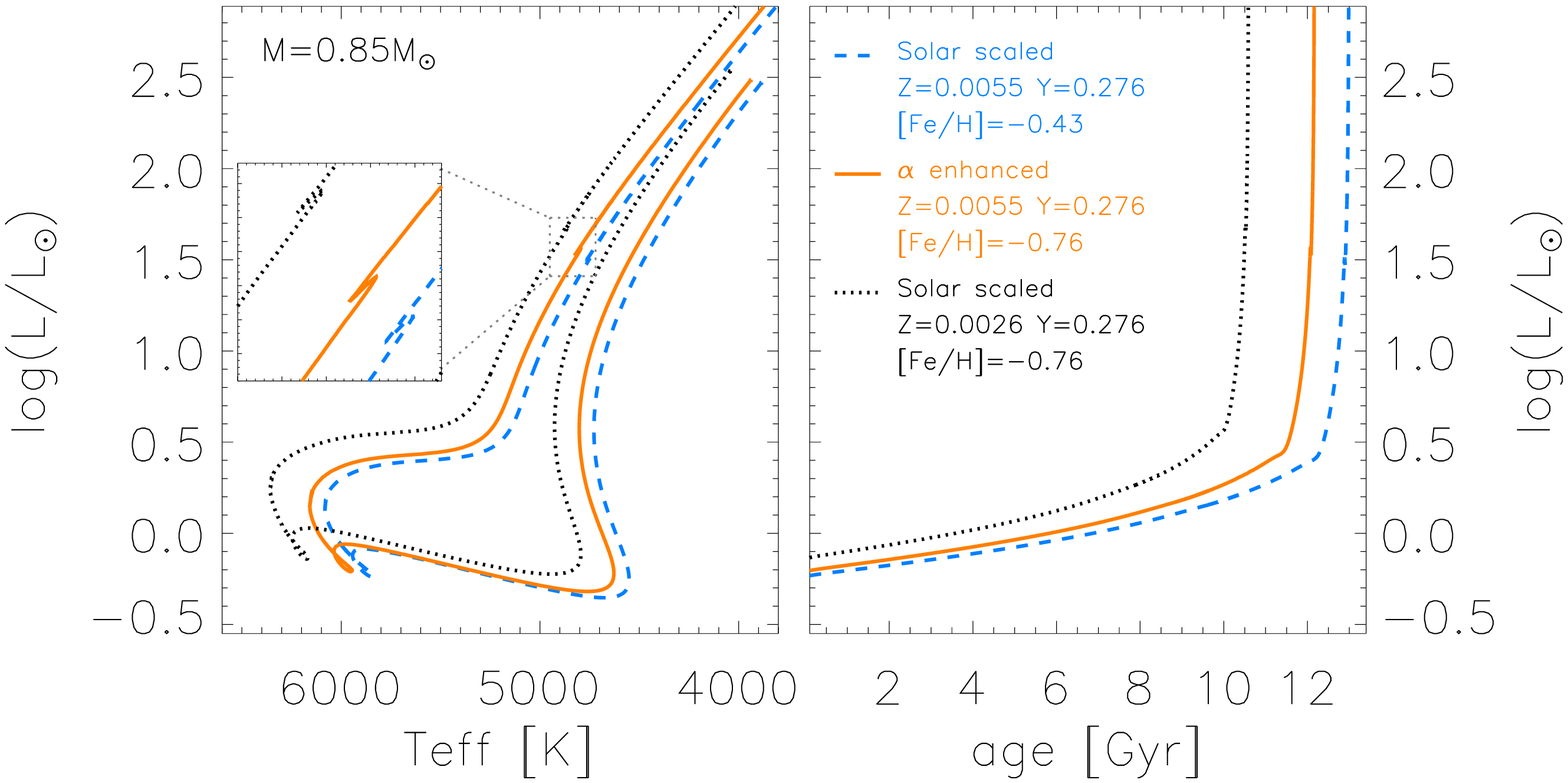}
              \caption{
              A comparison between the $\alpha$-enhanced evolutionary track (orange solid line)
              and the solar-scaled one with the same metallicity Z (blue dashed line). 
              For comparison,
              a solar-scaled evolutionary track
               with the same [Fe/H] value (black dotted line)
              is also displayed.
              The helium content and the stellar mass of the three stars are the same (Y=0.276,  M=0.85 \msun ).
              The left panel is HRD with sub-figure zoom-in around the red giant branch bump region.
              The right panel shows how the luminosity of the star evolve with time.}
              \label{fig:aph}
      \end{figure}

 It is worth noting here that when we change the heavy element number
 fractions ($N_i/N_Z$) to obtain a new metal partition in \parsec, their
 fractional abundances by mass ($Z_i/Z_{tot}$) are re-normalized in such
 a way that the global metallicity, $Z$, is kept constant.  Hence,
 compared to the solar partition at the same total metallicity Z, a model
 with enhanced $\alpha$-elements shows a depression of Fe and the related
 elements, because  the total metallicity remains unchanged by
 construction.  The  Hertzsprung-Russell diagram (HRD) in the left panel
 of   Fig. \ref{fig:aph} shows that, with the same total metallicity Z
 and helium content Y, the $\alpha$-enhanced star (orange solid line) is
 slightly hotter than the solar-scaled one (blue dashed line) both on the
 main sequence and on the red giant branch because of  the net effect of
 changes to the opacity.  Higher temperature  leads to a faster
 evolution, as illustrated in the right panel of Fig. \ref{fig:aph}.  It
 is also interesting to compare the $\alpha$-enhanced star to a
 solar-scaled one with the same [Fe/H] (black dotted line in  Fig.
 \ref{fig:aph} ).  With the same [Fe/H] but higher  total metallicity Z,
 the $\alpha$-enhanced star is cooler.  Indeed, \citet{VandenBerg2012}
 report that if keeping [Fe/H] constant, the giant branch is shifted to a
 cooler temperature with increased Mg or Si, while O, Ne, S abundances
 mainly affect the temperatures of main sequence and turn-off phases.

    Various initial helium abundance values, for a given metallicity, are  allowed in the new version of \parsec.
    In the previous versions the initial helium mass fraction of the stars was obtained from the helium-to-metals enrichment law:	
    \begin{equation}
          \label{eq:he}
          Y = Y_p + \frac{\Delta Y}{\Delta Z}Z ~=~  0.2485 + 1.78 \times Z
    \end{equation}
    where $Y_p$ is the primordial helium abundance \citep{Komatsu2011}, and
    $\Delta Y / \Delta Z$ is the helium-to-metal enrichment ratio.  Because of
    differences in the adopted primordial and solar calibration He and
    metallicity values by different authors, the above two parameters are
    slightly different in different stellar evolution codes.  The latest YY
    isochrone \citep{Spada2013} adopts the relation $Y=0.25+1.48Z$; DSEP
    \citep{Dotter2007a, Dotter2007, Dotter2008} uses $Y=0.245+1.54Z$; MIST
    \citep{Choi2016} gives $Y=0.249+1.5Z$, and BaSTI \citep{Pietrinferni2006}
    adopts $Y=0.245+1.4Z$.  However, observations reveal that the helium
    content does not always follow a single relation.  Differences in helium
    abundance have been widely confirmed in globular clusters  between stellar
    populations with very similar metallicity.  The evidence includes the
    direct He I measurement on blue horizontal branch star \citep[][for
    instance]{Villanova2009, Mucciarelli2014, Marino2014,  Gratton2015}, on
    giant stars \citep{Dupree2011, Pasquini2011}, and  the  splitting of
    sequences  in  colour-magnitude diagram (CMD) both of globular clusters in
    Milky Way \citep[e.g.][]{Bedin2004, Villanova2007, Piotto2007, Milone2008,
    DiCriscienzo2010} and in Magellanic Cloud clusters \citep{Milone2015,
    Milone2016}.  \citet{Bragaglia2010} found that the brightness of the RGB
    bump, which should increase with He abundance, is fainter in the first
    generation than the second generation in 14 globular clusters.  Indeed, He
    variation is considered one of the key parameters (and problems) to
    understand multiple populations in GCs \citep[see the review by ][and the
    references therein]{Gratton2012}.  In the new version of \parsec, we allow
    different helium contents at any given metallicity Z.

\section{Calibration with 47~Tuc}
\label{sec:47tuc}

Globular Clusters (GCs) have been traditionally considered as the paradigm of a
single stellar population, a coeval and chemically homogeneous population of
stars covering a broad range of evolutionary phases, from the low-mass main
sequence, to the horizontal branch (HB) and white dwarf sequences.  For this
reason they were considered the ideal laboratory to observationally study the
evolution of low mass stars and to check and calibrate the stellar evolution
theory. This picture has been challenged during the last two decades by
photometric and spectroscopic evidence of the presence of multiple populations
in most, if not all, globular clusters (for instance NGC 6397
\citep{gratton2001, milone20126397}, NGC 6752 \citep{gratton2001, milone2010},
NGC 1851 \citep{Carretta2014}, NGC 2808 \citep{Antona2005,Carretta2006,
Piotto2007, Milone20152808}, NGC 6388 \citep{Carretta2007}, NGC 6139
\citep{Bragaglia2015}, M22 \citep{Marino2011}, etc.).  Nevertheless, Globular
Clusters remain one of the basic workbenches for the stellar model builders,
besides their importance for dynamical studies and, given the discovery of
multiple populations, also for the early chemo-dynamical evolution of stellar
systems.

47~Tuc, a relatively metal-rich Galactic Globular Cluster, also shows evidence of the
presence of at least two different populations:
{\it i)} bimodality in the distribution of CN-weak and CN-strong targets, not
only in red giant stars \citep{Briley1997,norris79, harbeck03} but also in MS
members \citep{cannon98};
{\it ii)} luminosity dispersion in the sub-giant branch, low-main sequence and HB
\citep{Anderson2009,DiCriscienzo2010,Nataf2011,Salaris2016} indicating a dispersion in
He abundance;
{\it iii)} anti-correlation of Na-O in RGB and HB stars
\citep{Carretta2009,Carretta2013,Gratton2013}  and also in MS-TO ones
\citep{D'Orazi2010,Dobrovolskas2013}.

The presence of at least two different populations with different chemical
compositions seems irrefutable (even if their origin is still under debate).
Particularly convincing  is the photometric study by \citet[][and references
therein]{Milone2012}, which concludes, in good agreement with other works
\citep{Carretta2009, Carretta2013}, that for each evolutionary phase, from MS
to HB, the stellar content of 47~Tuc belongs to two different populations,
``first generation'' and ``second generation'' ones (thereafter, FG and SG
respectively).  The FG population represents $\sim 30$\%  of the stars, and it
is more uniformly spatially distributed than the SG population, which is more
concentrated in the central regions of the cluster.

Choosing 47~Tuc as a reference to calibrate PARSEC stellar models, requires
therefore computation of stellar models with metal mixtures corresponding to
the two identified populations.  In the next section we describe the sources to
derive the two different metal mixtures that will be used for the opacity and
EOS tables in the stellar model computations, and in the follow-up isochrone
fitting.

\subsection {Metal mixtures}
\label{subsec:met}

 Chemical element abundances are given in the literature as the absolute values
A(X)\footnote{$\rm{A(X)}=\log(\rm{N}_{\rm X}/\rm{N}_{\rm H}) + 12$, with ${\rm
N}_{\rm X}$ is the abundance in number for the element X.},
or as [X/Fe]\footnote{[X/Fe]=$\log({\rm N}_{\rm X}/{\rm N}_{\rm Fe})-\log({\rm
N}_{\rm X}/{\rm N}_{\rm Fe})_\odot$},
the abundance with respect to the iron
content and referred to the same quantity in the Sun. Since the solar metal
mixture has changed lately and since there is still a hot debate about the
chemical composition of the Sun, it is important to translate all the available
data to absolute abundances, taking into account the solar mixture considered
in each source.  We follow that procedure to derive the metal mixtures for the
first and second generation in 47~Tuc.

The separation between the two populations based on photometric colours done by
\citet{Milone2012} agree with the separation based on Na-O anti-correlation by
\citet{Carretta2009} and \citet{Gratton2013}.  We decide hence to use the same
criteria to classify the star as FG or SG member.

Concerning He mass fraction $Y$, the scatter in luminosity seen in some
evolutionary phases has been attributed to different amounts of He in the
stellar plasma (see references above).  The analyses presented in
\citet{Milone2012} suggests that the best fitting of the colour difference
between the two populations is obtained with a combination of different C, N
and O abundances, plus a small increase of He content in the SG ($\Delta
Y$=0.015-0.02).  These results agree with those presented in \citet[][$\Delta
Y$=0.02--0.03]{DiCriscienzo2010}, and rule out the possibility of explaining
the 47~Tuc CMD only with the variation of He abundance.

Table~\ref{tab:mix} lists the elemental abundances we adopt for the two
generations of 47~Tuc together with the corresponding references.  The
abundances of some elements, like carbon, nitrogen and oxygen,  may change
during the evolution because of standard (convection) and non-standard (i.e.
rotational mixing) transport processes.  Therefore,  CNO  abundances are
compiled from available measurements  for  MS/TO stars, and their sum
abundances are nearly the same in both populations.  Other elements  are not
expected to be affected by mixing processes during stellar evolution, so we use
the values measured mainly in the red giant phase where  hundreds of stars are
observed.  If available, abundance determinations that take into account NLTE
and 3D effects are adopted.  There is no clear abundance difference  of
elements Mg, Al, Si, Ca, and Ti between the two populations of 47~Tuc.  Since
their abundances show large scatter from different literature sources and are
sensitive to the choice of measured lines, we  use the mean values of the
literature abundances for both FG and SG.  The iron abundance [Fe/H]=$-$0.76
dex is adopted from \citet{Carretta2009feh} and \citet{Gratton2013} who measure
the largest giant sample and HB sample of 47~Tuc respectively with internal
fitting errors less than 0.02 dex.  The [Fe/H] values derived from giants are
much less dependent on the effects of microscopic diffusion than in the case of
main sequence stars.  We notice that in literature some authors suggest a
[Fe/H] dispersion of  $\sim0.1$ dex for 47~Tuc \citep{Alves-Brito2005}.  On the
other hand \citet{Anderson2009} conclude that for 47~Tuc a He dispersion of
$\sim0.026$ has an equal effect on the MS as a [Fe/H] dispersion of $\sim0.1$
dex.  Since we consider different He contents for the FG and SG stars, we do
not apply any further [Fe/H] dispersion.  Other elements which are not
displayed in table~\ref{tab:mix} keep the solar abundances ([X/Fe]=0).  The
anti-correlation between the abundances of C and N, as well as O and Na,
contributes to the main difference of the metal mixtures between FG and SG.
The difference in the final \alpfe values between the two generations is due to
the difference in O abundance.  The resulting metallicities are Z$_{\rm
FG}=0.0056$ for the first generation and Z$_{\rm SG}=0.0055$ for the second
generation, respectively.  Following \citet{Milone2012}, the assumed He
abundances are Y=0.256 and Y=0.276 for FG and SG, respectively.  Table
\ref{tab:par} lists the general metal mixture information for the two stellar
populations, including Z, Y, [M/H], [Fe/H], and \alpfe.  The referred solar
abundance is derived from \citet{Caffau2011}, as described in
\citet{Bressan2012}.  We consider eight $\alpha$ elements when calculating the
total $\alpha$~ enrichment \alpfe: O, Ne, Mg, Si, S, Ca, Ar, and Ti.  Thus for
the FG stars of 47~Tuc, [Z=0.0056, Y=0.256], \alpfe = 0.4057 dex and,  for the
SG stars, [Z=0.0055, Y=0.276], ~\alpfe= 0.2277 dex.  These two values,
approximately $\sim$ 0.4 and 0.2 respectively, are the typical \alpfe values
observed in $\alpha$-enriched stars.  Finally we note that we will adopt the
metal partitions of these two $\alpha$-enriched generations to calculate
stellar evolutionary tracks and isochrones also for other metallicities.

  \begin{table*}
          \caption{
                  Chemical element abundances  of 47~Tuc two stellar populations (FG and SG).
          The abundances are written in the format of [X/Fe],  their corresponding references  are also listed.
  }
          \centering
          \begin{tabular}{@{}lrrcc@{}}
                  \hline
                  & FG & SG & reference & note \\
                  \hline
                  $\mathrm{[C/Fe]}$        & 0.12 & -0.09& \citet{cannon98, Milone2012} & MS  \\
                  $\mathrm{[N/Fe]}$        & 0.32 & 1.17 & \citet{cannon98, Milone2012} & MS  \\
                  $\mathrm{[O/Fe]}$        & 0.42 & 0.17 & \citet{Dobrovolskas2013} & TO, NLTE+3D  \\
                  $\mathrm{[Ne/Fe]}$       & 0.40 & 0.40 & ----                     & estimated  \\
                  $\mathrm{[Na/Fe]}$       & -0.12& 0.10  & \citet{Dobrovolskas2013} & TO, NLTE+3D  \\
                  $\mathrm{[Mg/Fe]}$       & 0.32 & 0.32 & \citet{Carretta2009,Carretta2013,Gratton2013,Cordero2014,Thygesen2014} & mean value of RGB/HB  \\
                  $\mathrm{[Al/Fe]}$       & 0.20 & 0.20 & \citet{Cordero2014, Thygesen2014} & mean value of RGB  \\
                  $\mathrm{[Si/Fe]}$       & 0.27 & 0.27 & \citet{Gratton2013, Thygesen2014, Carretta2009, Cordero2014} & mean value of RGB/HB  \\
                  $\mathrm{[S/Fe]}$        & 0.40 & 0.40 & ---- & estimated  \\
                  $\mathrm{[Ca/Fe]}$       & 0.27 & 0.27 & \citet{Carretta2009,Gratton2013,Cordero2014,Thygesen2014} & mean value of RGB/HB \\
                  $\mathrm{[Ti/Fe]}$       & 0.20 & 0.20 & \citet{Cordero2014, Thygesen2014} & mean value of RGB/HB \\
                  \hline
          \end{tabular}
          \label{tab:mix}
  \end{table*}

  \begin{table}
          \caption{General metal mixture of 47~Tuc two stellar populations (FG and SG).}
          \centering
          \begin{tabular}{@{}crr@{}}
                  \hline
                  & FG & SG \\
                  \hline
                  Z  & 0.0056   & 0.0055    \\
                  Y  & 0.256    & 0.276 \\
                  {[M/H]} & $-$0.43  & $-$0.41    \\
                  {[Fe/H]} & $-$0.76  & $-$0.76   \\
                  \alpfe$^\dagger$  &  0.41   & 0.23    \\
             \hline
          \end{tabular}
          \label{tab:par} \\
          $^\dagger$~ Labeled as \alpfe $\sim$ 0.4 and 0.2. Difference in the \alpfe values is due to O abundance differences. \\
  \end{table}

\subsection{Isochrones fitting and Luminosity function}
\label{subsec:fitting}

    With the detailed metal mixture and helium abundance of 47~Tuc, we
    calculate new sets of evolutionary tracks and isochrones, and transform
    them into the observational color-magnitude diagram (CMD) in order to fit
    the data.  This fitting procedure, based on our adopted model prescriptions
    (e.g. mixing length, atmospheric boundary condition, bolometric
    corrections), aims to calibrate other  parameters (e.g. extra mixing) in
    the model as described below.

  \subsubsection{Low main sequence to turn-off}
    \label{sub:ms}

      \begin{figure}
              \centering
              \includegraphics[width=0.5\textwidth,angle=0]{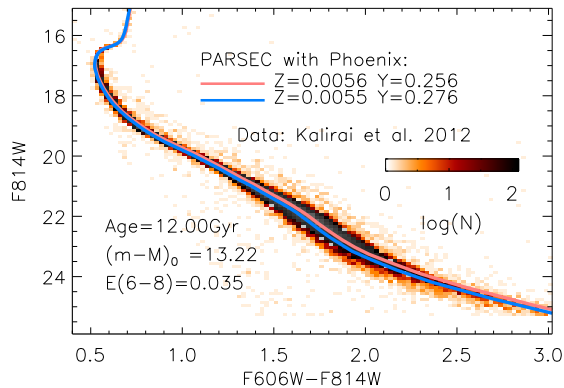}
              \caption{Isochrone fitting with Hess diagram of 47~Tuc data for the low main sequence\citep{kalirai}.
              The bin size of Hess diagram is 0.025 mag in color and 0.1 mag in F814W magnitude.
              The fitting parameters (age, $\eta$, (m$-$M)$_0$, and E(6$-$8)) are listed in the legend.}
              \label{fig:fitk}
      \end{figure}

      \citet{kalirai} provides deep images of 47~Tuc taken with the Advanced
      Camera for Surveys (ACS) on \textit{Hubble Space Telescope} (HST). The
      corresponding colour-magnitude diagrams cover the whole main sequence of
      this cluster, till the faintest stars.  Fig.  \ref{fig:fitk} shows our
      isochrone fitting of their photometric data, i.e. in the F606W and F814W
      bands.  In order to display the  relative density of stars on CMD the
      data are plotted with the Hess diagram (binsize 0.025 mag in color and
      0.1 mag in I magnitude).  By assuming a standard extinction law
      \citep{Cardelli1989} we derive, from the isochrone fitting, an age of
      12.00~Gyr,  a distance modulus  (m$-$M)$_0$ of 13.22 ( (m$-$M)$_{F606W}
      \sim$13.32),  and a reddening of E(F606W-F814W)=0.035 (hereafter named
      E(6-8)).  The fitting is performed not only to the main sequence but also
      to the giant branch and HB phase as we will show later in
      Sec.~\ref{sub:rgb} and Sec.~\ref{subsec:hb}.  The values of age,
      (m$-$M)$_0$, and E(6-8)  are adjusted with visual inspection
      with the  priority of
      improving the turn-off and HB fittings.

    The distance modulus of 47~Tuc has been determined by many other authors,
    however with different results.  For instance, using  HST proper motion
    \citet{watkins2015} derive a distance of 4.15 kpc ((m$-$M)$_0\sim$13.09)
    which is lower than the values in the Harris catalog \citep[4.5 kpc
    ((m$-$M)$_0\sim$13.27) and (m$-$M)$_V$=13.37,][2010 edition]{Harris1996};
    the eclipsing binary distance measurement \citep[(m$-$M)$_V$=
    13.35,][]{Thompson2010}; the result based on the  white dwarf cooling
    sequence \citep[(m$-$M)$_0\sim$13.32,][]{Hansen2013}; and that derived from
    isochrone fitting to BVI photometry
    \citep[(m$-$M)$_V$=13.375,][]{Bergbusch2009}.  Our best fit distance lies
    in between them, and agrees with other recent distance modulus
    determinations \citep[e.g. ][(m$-$M)$_0$=13.21$\pm$0.06 based on the
    eclipsing binary]{Brogaard2017}.  \gaia will release the  parallaxes and
    proper motions including stars in 47~Tuc in its DR2 in early 2018, and will
    help to solve the distance problem.  However, we will show in the following
    section that our best estimate result offers a very good global fitting,
    from the very low main sequence till the red giant and horizontal branches.

   \subsubsection{RGB bump and envelope overshooting calibration}
   \label{subsuc:eov}

Some  GC features in CMD are very sensitive to stellar model parameters which
are, otherwise, hardly constrained from observations directly.  This is the
case of the efficiency of mixing below the convective envelope ( envelope
overshooting), that is known to affect the luminosity of the red giant branch
bump (RGBB).  In this section we will use the 47~Tuc data to calibrate the
envelope overshooting to be used in low mass stars by \parsec.

The RGB bump is one of the most intriguing features in the CMD.  When a star
evolves to the ``first dredge-up'' in the red giant phase, its  surface
convective zone deepens while the burning hydrogen shell moves outwards.  When
the hydrogen burning shell encounters the chemical composition discontinuity
left by the previous penetration of the convective zone, the sudden increase of
H affects the efficiency of the burning shell and the star becomes temporarily
fainter.  Soon after a new equilibrium is reached, the luminosity of the star
raises  again.  Since the evolutionary track crosses the same luminosity three
times in a short time, there is an excess of star counts in a small range of
magnitudes, making a ``bump'' in the star number distribution (Luminosity
Function) along the red giant branch.  This is because the number of stars in
the post main sequence phases is proportional to the evolutionary time of the
stars in these phases. The longer the crossing time  of the chemical
composition discontinuity by the burning shell, the more the stars accumulate
in that region of the RGB.

    The properties of RGBB, including the brightness and the extent,
    are important to study the stellar structure and to investigate the nature of GCs.
    47~Tuc was the first GC where the existence of the RGBB was  confirmed~\citep{king1985}.
    Since then, many works, both theoretical and observational
    \citep[for instance,][]{Alongi1991, Cassisi1997, Zoccali1999, Bono2001, Cassisi2002,
    Bjork2006, Salaris2006, Cecco2010, Bragaglia2010, Cassisi2011,  Nataf2013},
    have studied the features of RGBB.	

    The intrinsic brightness and extent of the RGBB are sensitive to:

    \begin{description}
            \item[ Total metallicity and metal partition:]

                    \citet{Nataf2013} propose an empirical function of RGBB extent to metallicity:
                    the more metal-poor the globular cluster is,    the smaller is the extent of the RGBB.	
                    From the theoretical point of view, stars with lower total metallicity are brighter 
                    compared to the higher metallicity stars,
                    causing their hydrogen burning shell to move outwards faster.
                    Since they are also hotter, the surface convective envelope is thinner 
                    and the chemical composition discontinuity is smaller and less deep.
                    As a consequence,  their RGBB is very brief and covers a small range of magnitudes at higher luminosity.
                    This is why RGBB in metal-poor globular clusters is very difficult to be well sampled.
                    The metal partition also affects the features of RGBB, even with the total metallicity remains the same.
                    As already shown in Fig. \ref{fig:aph} and in Sec. \ref{sec:input},
                    a stellar track with $\alpha$-enhancement is hotter than the solar-scaled one 
                    with the same total metallicity Z because of different opacity,
                    leading to a brighter RGBB.
                    Since CNO are the most affected elements in the giant branch
                    and they are important contributors to the opacities, 
                    their varying abundances have an important impact on the location of RGBB.
                    e.g. \citet{Rood1985} shows that
                    enhancing CNO by a factor of ten has
                    larger effect on the RGBB luminosity 
                    than enhancing Fe by a factor of ten over the same metallicity Z.
                    More recently, \citet{VandenBerg2013b} show that higher oxygen abundance leads to a fainter RGBB
                    if the [Fe/H] is fixed.

            \item[Helium content:]	

                    A larger helium content renders the  star hotter and brighter \citep{Fagotto1994}.
                    \citet{Bragaglia2010} studied the RGBB of 14 globular clusters and found that
                    the more He-rich second generation shows brighter RGBB than the first generation.
                    Similar to the mechanism in metal-poor stars,
                    hot He-rich stars have less deep convective envelopes 
                    and their high luminosity makes the hydrogen burning shells to move faster across the discontinuity.
                    Hence with the same total metallicity Z and stellar mass
                    the RGBB of the He-rich star is brighter, 
                    less extended, 
                    and more brief. 
                    In table \ref{tab:hebump}
                    we take a M=0.85 \msun~ star as an example to show 
                    how the RGBB luminosity and evolution time vary
                    with different helium contents.

                    \begin{table*}
                            \caption{
                            RGBB parameters of stars with a constant mass (M=0.85 \msun)
                            and  metallicity (Z=0.0055) but different helium contents.
                            Mean luminosity ~$\bar{log(L/L_\odot)}_{RGBB}$ ,
                            luminosity extent~$\Delta$ log(L)$_{RGBB}$,
                            RGBB beginning time t$_{0,RGBB}$,
                            and RGBB  lifetime ~$\Delta$ t$_{RGBB}$ are listed. }
                            \centering
                            \label{tab:hebump}
                            \begin{tabular}{@{}cccccc@{}}
                                    \hline
                                    Z  &  Y & $\bar{log(L/L_\odot)}_{RGBB}$  & $\Delta$ log(L)$_{RGBB}$ & t$_{0,RGBB}$~(Gyr)& $\Delta$ t$_{RGBB}$~(Myr)\\
                                    \hline
                                    0.0055 & 0.276 &1.5443  & 0.03557 &12.062	& 27.159\\
                                    0.0055 & 0.296 &1.5887  & 0.03177 &10.584	& 22.637\\
                                    \hline
                            \end{tabular}
                    \end{table*}

            \item[Age:]

                    Stars with younger age are hotter,  with their thinner convective envelope their RGBB are brighter.
                    In principle multiple populations born in different ages spread the GC RGBB luminosity.
                    However,  considering that the age variation of the multiple populations is usually small ($\sim$ a few Myr),
                    it contributes little to the GC RGBB luminosity spread  compared to the He variation
                    \citep[see, e.g.][]{Nataf2011tuc}.

            \item[Mixing efficiency:]

        The mixing efficiency of the star, both mixing length  and envelope
        overshooting (EOV), determines the maximum depth of the convective
        envelope and affects the brightness and evolutionary time of RGBB.  The
        more efficient the mixing is, the deeper the convective envelope is,
        the earlier the hydrogen-burning shell meets the discontinuity left by
        the penetration of the surface convective zone and the fainter the RGBB
        is.	 		For the mixing length, we adopt the solar-calibrated
        value~$\alpha_{MLT}^{\odot}$=1.74 in \parsec as described in
        \citep{Bressan2012}.  The EOV is calibrated with the new stellar tracks
        against the observations of the RGBB of 47~Tuc.
\end{description}

        Overshooting is the non-local mixing that may occur at the borders of any convectively unstable region
         \citep[i.e.,][and references therein]{bressan14}.
        The extent of the overshooting at the base of the convective envelope 
        is called envelope overshooting,
        and the one above the stellar convective core is called core overshooting.
        There are observations that can be better explained with envelope overshooting, 
        for instance, the blue loops of intermediate and massive stars~\citep{Alongi1991, Tang2014},
        and the carbon stars luminosity functions in the Magellanic Clouds, that require a more
        efficient third dredge-up in AGB stars~\citep{Herwig00, Marigo07}.
        At the base of the convective envelope of the Sun,
        models with an envelope overshooting  of $\Lambda_e \approx~ 0.3 \sim 0.5~H_p$
        (where $H_p$~is the pressure scale height)
        provide a better agreement with the helioseismology data~\citep{Christensen-Dalsgaard2011}.
        The envelope overshooting also affects   the surface abundance of light elements \citep{pms},
        and asteroseismic signatures in stars \citep{LebretonGoupil2012}.
          
          \begin{figure}
                  \centering
                  \includegraphics[width=.48\textwidth,angle=0]{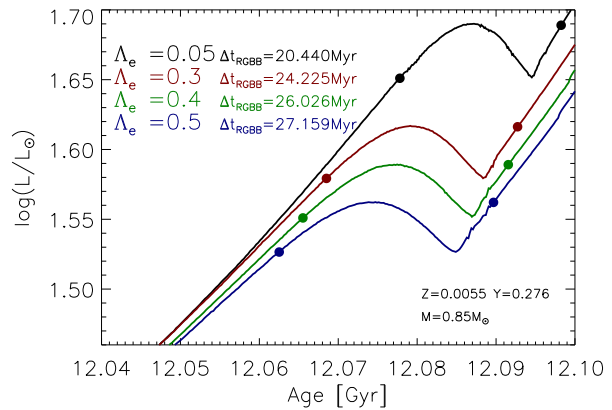}
                  \caption{The RGBB luminosity as a function of stellar age for a 0.85 \msun~ star  but with different EOV.
                  The black, red, green, and blue line from top to bottom represent tracks with~$\Lambda_e$ = 0.05, 0.3, 0.4, and 0.5.
                  The filled dots mark the minimum and maximum luminosity of RGBB for each track,
                  and ~$\Delta t_{RGBB}$~ is the evolution time from the  minimum luminosity to the maximum one.}
                  \label{fig:btime}
          \end{figure}		

        In Fig.  \ref{fig:btime}~ we compare the RGBB evolution of models computed with different EOV values, $\Lambda_e$,
        at the same stellar mass and composition.
        Every pair of filled dots marks the brightness extent of the RGBB.
        The figure shows that a larger envelope overshooting not only makes the RGBB fainter,
        but also of longer duration, leading to a  more populated RGBB.	
        A larger EOV value leads to a deeper surface convective zone,
        and the hydrogen burning shell encounters the chemical discontinuity earlier.	    	

        \begin{figure}
                \centering
                \includegraphics[height=0.25\textheight]{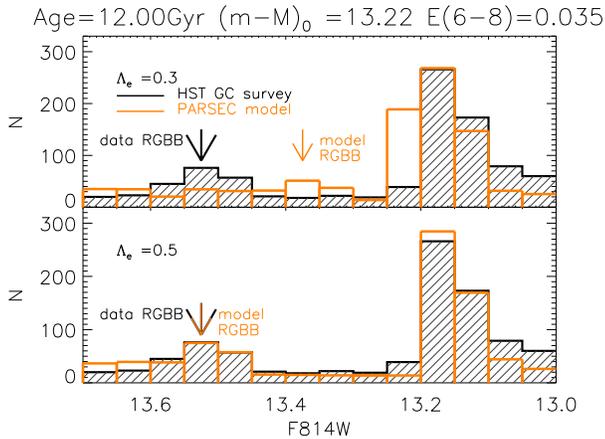}
                \caption{Comparison between LF of 47~Tuc data \citep{acsgc}~and the new \parsec isochrone with different EOV.
                The fitting parameters (age, (m-M)$_0$, and E(6-8))  the same as in Fig.\ref{fig:fitk}.
                The black histogram filled with oblique lines is the data LF,
                whilst orange histogram is LF derived from new \parsec isochrones with 
                30\% contribution from the FG of 47~Tuc and 70\% from the SG.
                The upper panel isochrones of each sub-figure  are calculated with EOV value $\Lambda_e$=0.3,
                and the lower panel are the ones with $\Lambda_e$=0.5.
                Orange arrow and black arrow mark the location of RGBB in model and in data, respectively.
                The bin size of the LF is 0.05 mag.}
                \label{fig:lfcom}
        \end{figure}

        The Luminosity Function (LF) is a useful tool to compare the observed
        morphology of RGBB with that predicted by the theory.  Taking into
        account that the 47~Tuc population contribution is 30\% from the FG and
        70\% from the SG as suggested by \citet{Milone2012} and
        \citet{Carretta2009}, we simulated the LF of 47~Tuc with our isochrones
        with different EOV values.  The comparison between the observed and
        predicted LFs is shown in  Fig.  \ref{fig:lfcom}.  For the observed LF
        we have used data from the HST/ACS survey of globular clusters
        \citep{acsgc}.  Both observations and models are sampled in bins of
        0.05 magnitudes.  The fitting parameters are the same as those we used
        in Fig.~\ref{fig:fitk}.  The model LF (orange histogram) are calculated
        with envelope overshooting $\Lambda_e$=0.3 in the upper panel and with
        $\Lambda_e$=0.5 in the lower panel.	It is evident that the LF computed
        with the small envelope overshooting value $\Lambda_e$=0.3 has RGBB too
        bright compared to data (black histogram filled with oblique lines).
        We find that the agreement between observations and models is reached
        when one adopts  a value of $\Lambda_e=0.5 H_p$ below the convective
        border, with our adopted metal mixtures and best fit isochrone.  This
        provides a robust calibration of the envelope overshooting parameter.
        This envelope overshooting calibration will be applied to all other
        stellar evolution calculations of low mass stars.

   \subsubsection{Red Giant Branch}
    \label{sub:rgb}

\begin{figure}
            \centering
            \includegraphics[width=.48\textwidth,angle=0]{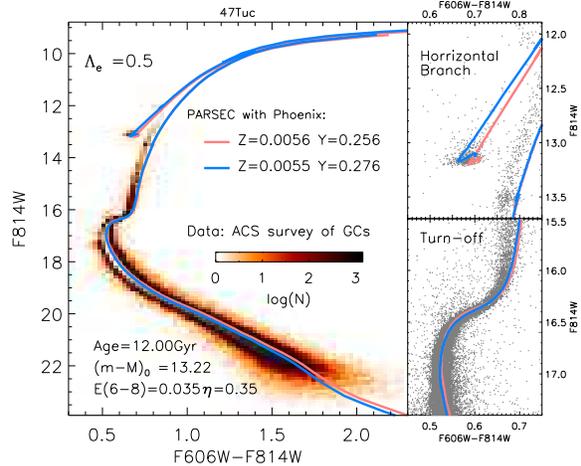}
            \caption{Isochrone fitting with Hess diagram (the left panel) of 47~Tuc data \citep{acsgc} for all the evolutionary phases,
            and with scatter plots highlighting  the horizontal branch region (the upper right panel)
            and the turn-off region (the lower right panel).
            The red line and blue line represent isochrones of the first and second generation, respectively,
            as the legend shows.
            The fitting parameters are:
            age=12.00 Gyr, (m$-$M)$_0$=13.22, E(6-8)=0.035. }
            \label{fig:fitp2}
    \end{figure}

    While \citet{kalirai} focus on the faint part of the main sequence as shown
    in Fig. \ref{fig:fitk}, another dataset of 47~Tuc, the HST/ACS survey of
    globular clusters \citep{acsgc}, is devoted to the  Horizontal Branch
    \citep{Anderson2008} with the same instrument.  In Fig.  \ref{fig:fitp2} we
    show the global fitting of ACS data of 47~Tuc, from main sequence up to the
    red giant branch and HB.  The best fitting parameters we derived are the
    same as those we used to fit the lower main sequence data in Fig.
    \ref{fig:fitk}.  The Hess diagram is used for the global fitting (the left
    panel) with bin size 0.025 mag in color and 0.1 mag in F814W magnitude.	The
    HB region and the turn-off region are zoomed in with scatter plots in the
    two right panels.  Thanks to the detailed composition derived from the
    already quoted observations for the two main populations of 47~Tuc and the
    new computed models, by assuming a standard extinction law
    \citep{Cardelli1989} and using the adopted EOV and mixing length
    parameters, we are able to perform a global fit to the CMD of 47~Tuc
    covering almost every evolutionary phase over a range of about 13
    Magnitudes.  This must be compared with other fittings that can be found in
    literature and that usually are restricted to only selected  evolutionary
    phases \citep[][etc.]{Kim2002, Salaris2007, VandenBerg2013, VandenBerg2014,
    Chen2014, McDonald2015}.  However, it is worth  noting that the distance of
    this cluster, as already discussed in Sec.~\ref{sub:ms}, together with the
    cluster age, have varied over the years in many careful studies.  We look
    forward to \gaia~ DR2 to put more constrains to this problem.

  As the upper right panel of Fig.~\ref{fig:fitp2} show, the isochrones
  corresponding to both of the two stellar generations run on the red side of
  the data in the RGB phase.  Part of the discrepancy could be explained by the
  bolometric correction used.  Here we are using bolometric correction from
  \texttt{PHOENIX} atmosphere models as described in \citet{Chen2015} for
  \parsec \texttt{v1.2S}, where only the total metallicity is considered in the
  transformation of log(L) vs. log(Teff) into F814W vs. (F606W-F814W).  As the
  metallicities of the two 47~Tuc populations (Z=0.0056 and Z=0.0055) show only
  a marginal difference, we adopt for the two populations the same bolometric
  corrections.  Thus Fig.~\ref{fig:fitp2} reflects basically the difference of
  the two populations in the theoretical $\log(L)$ vs. $\log(T_{\rm eff})$ HR
  diagram.  This ``RGB-too-red'' problem also exists in \citet{Dotter2007},
  when they fit the same set of data using DSEP models (see their Figure   12),
  as they apply bolometric correction from \texttt{PHOENIX} as well.

\begin{figure}
            \centering
            \includegraphics[width=.48\textwidth,angle=0]{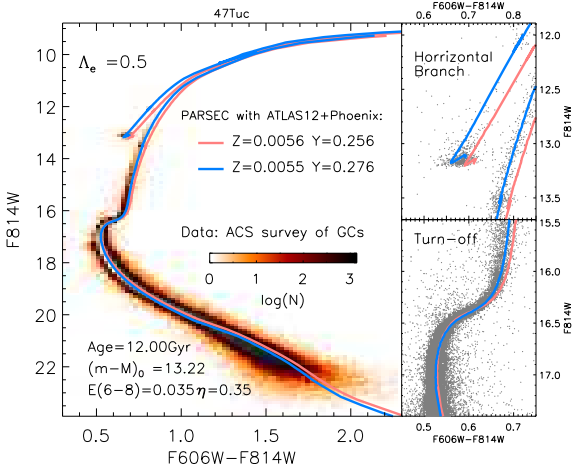}
            \caption{The same isochrone fitting with Hess diagram and scatter plots of 47~Tuc data \citep{acsgc} as in Fig.~\ref{fig:fitp2},
            but with atmosphere models from ATLAS12 for Teff hotter than 4000~K.}
            \label{fig:fita12}
    \end{figure}

  To minimize this discrepancy,   we use the ATLAS12 code \citep{Kurucz2005},
  which considers not only the total metallicity Z but also log(g) and detailed
  chemical compositions for the color transformation, to compute new atmosphere
  models with our best estimate chemical compositions of the two 47~Tuc
  populations.  We adopt these ATLAS12 models for the new fits to 47~Tuc, but
  only for models  with Teff hotter than 4000 K ((F606W-F814W)$\sim$1.3).  For
  lower Teff we still use \texttt{PHOENIX} because ATLAS12 models may be not
  reliable at cooler temperatures \citep{Chen2014}.  Here we show the fit
  obtained with ATLAS12+\texttt{PHOENIX} bolometric correction  in
  Fig.~\ref{fig:fita12}.  We see that with the same fitting parameters as in
  Fig.   \ref{fig:fitp2}, the prediction of the RGB colors is improved by
  applying new ATLAS12 bolometric correction.  The two stellar generation are
  split on RGB phase in Fig.   \ref{fig:fita12}.  We see that the SG
  (Z=0.0055), which is the main contributor as suggested by \citet{Milone2012}
  and \citet{Carretta2009}, is consistent with the denser region of the RGB
  data.  In other evolutionary phases the new ATLAS12 bolometric corrections do
  not bring noticeable changes.

  Since ATLAS12 only slightly affects the color of the RGB base, and the
  remainder of this paper deals with the LF of the bump and of the HB, in the
  following discussion, we will continue to use the standard atmosphere models
  of \parsec \texttt{v1.2S}.  PARSEC isochrones with ATLAS12 atmosphere models
  will be discussed in detail in another following work (Chen et al. in prep.)

\begin{figure}
            \centering
            \includegraphics[width=.48\textwidth,angle=0]{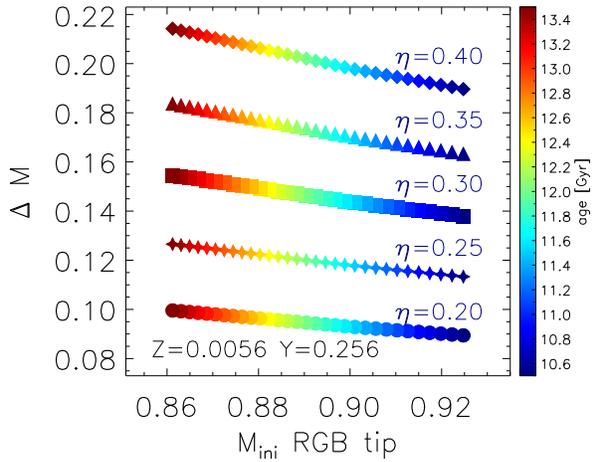}
            \caption{RGB mass lost in unit of \msun~ for FG of 47~Tuc (Z=0.0056, Y=0.256).
            The X axis is the initial mass of the tip  RGB star,
            and the Y axis shows the mass lost in this star during RGB phase.
            Five different  efficiency factors $\eta$ are illustrated,
            from top to bottom $\eta$=0.40 (filled diamond),  $\eta$=0.35 (filled triangle),
            $\eta$=0.30 (filled square),  $\eta$=0.25 (filled star), and  $\eta$=0.20 (filled dots).
            The color code displays the age, as shown in the color bar.}
            \label{fig:mlfg}
    \end{figure}

         Mass loss by stellar winds during the RGB phase has been considered
         for low mass stars, using the empirical formula by \citet{Reimers1975}
         multiplied by an efficiency factor $\eta$.  In Fig.~\ref{fig:mlfg} we
         show the mass lost by RGB stars in unit of \msun~ for the FG of 47~Tuc
         (the plot for SG is very similar).  Different efficiency factors
         ($\eta$) and ages are applied.  $\Delta M$ in the figure is the
         difference between the initial mass and current mass of the tip RGB
         star: $\Delta M=M_{initial}-M_{current}$.  The lost mass, which is
         greater with larger $\eta$, is an increasing function of the cluster
         age.  It is very difficult to derive observationally the mass lost in
         RGB stars directly since an accurate mass is not easy to derive and
         the RGB tip  is hard to identify.  However,  the RGB mass loss
         characterises the HB morphology, and this will be discussed  in next
         section.

    \subsubsection{Horizontal Branch morphology}
     \label{subsec:hb}

    The morphology of the Horizontal Branch in globular clusters is widely
    studied since the ``second parameter problem'' \citep[that is, the colour
    of the HB is determined not only by metallicity,][] {Bergh67, Sandage67}
    was introduced.  Aside from metallicity as the ``first parameter'', age ,
    He content, mass-loss, and cluster central density have been suggested as
    candidates to be the second, or even third, parameter affecting the
    morphology of the HB \citep[][.etc.]{Fusi97, Catelan2008, Dotter2010,
    Gratton2010, McDonald2015, Dantona2002, CaloiDantona05}.    Most of these
    parameters involve an effect on the mass of the stars which populate the
    cluster HB.  Stars with smaller stellar mass are hotter in temperature and
    bluer in color.  The HB stellar mass decreases as the cluster ages.  At a
    given age, He-rich star evolves faster and reach the Zero-Age Horizontal
    Branch (ZAHB) with lower mass.  If the age and He content are the same, the
    mass of HB stars is fixed by the mass loss along the RGB (here the mass
    loss driven by the helium flash is not considered).  Although the RGB mass
    loss  does not significantly affect the  RGB evolutionary tracks, it
    determines the location of the stars on the HB, by tuning the stellar mass.
    Here we illustrate how helium content and the RGB mass loss affect the HB
    morphology in the case of 47~Tuc.	

\begin{figure*}
              \centering
              \includegraphics[width=0.99\textwidth,angle=0]{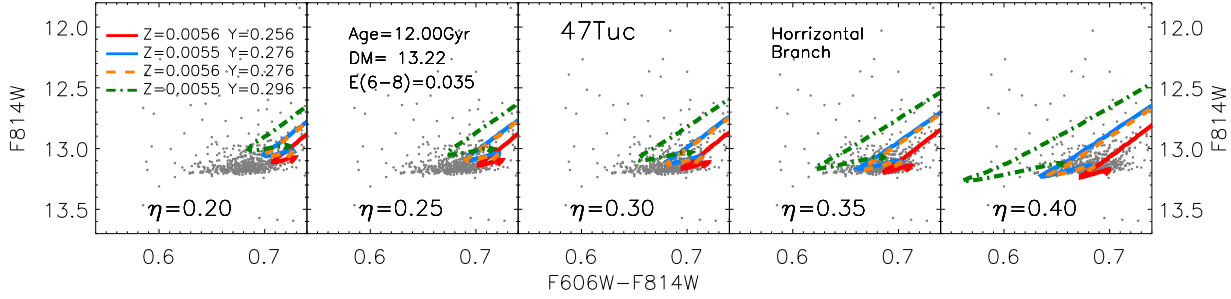}
              \caption{Horizontal branch morphology for different RGB mass loss parameters ($\eta$)~and metal/helium abundances,
              with the same isochrone fitting parameters (age, (m$-$M)$_0$, and E(6-8)) as in Fig.   \ref{fig:fitp2}.
              The red solid line, blue solid line, orange dashed line, and green dash-dot line
              represent isochrones of [Z=0.0056, Y=0.256],~[Z=0.0055, Y=0.276],~[Z=0.0056, Y=0.276], and [Z=0.0055, Y=0.296], respectively.
              The mass lost during the RGB  in  unit of \msun ~for each ~$\eta$ ~and metal/helium abundance is listed in Table \ref{tab:hehb2}.    }
              \label{fig:hb2}
      \end{figure*}

      The HB morphology with five different values of $\eta$ is displayed in
      Fig.~\ref{fig:hb2} for our best fitting parameters  derived in
      Sec.~\ref{sub:rgb}.  Different metal/helium abundances ([Z=0.0056,
      Y=0.256],~[Z=0.0055, Y=0.276],~[Z=0.0056, Y=0.276], and
      [Z=0.0055~Y=0.296]) are displayed.  The isochrones with Z=0.0056 are
      calculated with \alpfe $\sim$0.4 and those with Z=0.0055 are calculated
      with \alpfe $\sim$0.2.  The 47~Tuc data \citep{acsgc} are also plotted
      for comparison.  The differences between the isochrone with [Z=0.0055,
      Y=0.276] (blue solid line) and the one with [Z=0.0056, Y=0.276] (orange
      dashed line) are negligible on the HB, even though they refer to a
      different $\alpha$-enhanced mixture.  With the same RGB mass loss factor
      $\eta$, He-rich stars have their HB more extended (because of smaller
      stellar mass), bluer (due to both the smaller stellar  mass and the
      He-rich effect on radiative opacity), and more luminous  (because of
      larger He content in the envelope).  For stars with larger mass loss
      efficiency $\eta$ during their RGB phase, their HB is bluer, fainter, and
      more extended, because of smaller stellar mass (hence smaller envelope
      mass, since the core mass does not vary significantly with the mass loss
      rate).  Indeed, the effects of a higher He content and of a  lower mass
      (no matter if it is the result of an older age or a larger RGB mass loss)
      on HB stars are difficult to distinguished by means of the color, but can
      be disentangled because the larger helium content makes the He-rich star
      slightly more luminous.

      Table \ref{tab:hehb2} lists the current mass, $M_{ZAHB}$, of the first HB
      star and the corresponding mass that has been lost  $\Delta M_{RGB}$, in
      unit of \msun~.  In the table we also show the HB mass range $\delta
      M_{HB}$  that produces the corresponding color extent of HB.  All cases
      displayed in  Fig.   \ref{fig:hb2} are itemized.
      
      If one considers a uniform mass loss parameter $\eta$ for the two
      populations of 47~Tuc ([Z=0.0056, Y=0.256], and [Z=0.0055, Y=0.276]),
      $\eta=0.35$ is the value that  fits better the HB morphology in our best
      fitting case, as  Fig.~\ref{fig:hb2} illustrates.  As shown in  Table
      ~\ref{tab:hehb2}, a RGB mass loss parameter of $\eta=0.35$  leads to a
      value of the mass lost during RGB  between 0.1737 \msun \textendash
      0.1755 \msun.  In the literature there is a discrepancy among the results
      on RGB mass loss derived with different approaches, namely:  cluster
      dynamics, infrared excess due to dust, and HB modelling for this cluster.
      \citet{Heyl2015} study the dynamics of white dwarf in 47~Tuc, and
      concluded that the mass lost by stars at the end of the RGB phase should
      be less than about 0.2 \msun.  \citet{Origlia2007} observe the
      circumstellar envelopes around RGB stars in this cluster from mid-IR
      photometry and find the total mass lost on the RGB is $\approx 0.23 \pm
      0.07$~\msun.  \citet{McDonald2015} use HB star mass from literature to
      study the RGB mass loss and derive a Reimers factor $\eta=0.452$
      (corresponding to a RGB mass loss greater than $\sim$ 0.20 \msun).  Most
      recently, \citet{Salaris2016} assume  a distribution of the initial  He
      abundance to simulate the observed HB of 47~Tuc. They derive a lower
      limit to the RGB mass loss  of about 0.17 \msun,  but larger values are
      also possible, up to 0.30 \msun, with younger age, higher metallicity and
      reddening.  Our RGB mass loss results, based on a uniform mass loss
      parameter and our best fitting case of the two populations of this
      cluster, is consistent with the lower and upper limit values from the
      literature.  However, the real situations of the RGB mass loss in GCs, as
      discussed in the references above, could be much more complicated.  In
      our final database of the new \parsec isochrones we will provide
      different choices of He contents and mass loss parameters for the users'
      science purpose.

    \begin{table}
            \centering
            \caption{The mass lost during the RGB in  unit of~\msun~for different $\eta$ and metal/helium abundance.
            The current mass of the first HB star is  $M_{ZAHB}$,
            and $\Delta M^{RGB}$  represents its RGB mass loss in unit of \msun.
            The HB mass range is itemized in the last column $\delta M_{HB}$.
            All values listed here are derived from isochrones with
            age=12.0 Gyr, (m$-$M)$_0$=13.22, and E(V$-$I)=0.035,  as shown on Fig.  \ref{fig:hb2}.}
            \begin{tabular}{cccccc}
                    \hline \hline
                    Z & Y & $\eta$ &  $M_{ZAHB}$ (\msun) & $\Delta M_{RGB}$  (\msun) & $\delta M_{HB}$ (\msun) \\
                    \hline
                    0.0056 &    0.256   &  0.20 &   0.795832   &  0.0946    & 0.0023	\\
                    &       &  0.25 &   0.770375   &  0.1201    & 0.0033	\\
                    &       &  0.30 &   0.744052   &  0.1464    & 0.0044	\\
                    &       &  0.35 &   0.716765   &  0.1737    & 0.0053	\\
                    &       &  0.40 &   0.688402   &  0.2020    & 0.0059	\\
                    0.0055 &    0.276   &  0.20 &   0.758027   &  0.0953    & 0.0029	\\
                    &       &  0.25 &   0.732270   &  0.1211    & 0.0039	\\
                    &       &  0.30 &   0.705582   &  0.1478    & 0.0051	\\
                    &       &  0.35 &   0.677852   &  0.1755    & 0.0061	\\
                    &       &  0.40 &   0.648949   &  0.2044    & 0.0067	\\
                    0.0056 &    0.276   &  0.20 &   0.763649   &  0.0948    & 0.0028	\\
                    &       &  0.25 &   0.738049   &  0.1204    & 0.0039	\\
                    &       &  0.30 &   0.711533   &  0.1470    & 0.0050	\\
                    &       &  0.35 &   0.683996   &  0.1745    & 0.0059	\\
                    &       &  0.40 &   0.655309   &  0.2032    & 0.0066	\\
                    0.0055 &    0.296   &  0.20 &   0.726732   &  0.0954    & 0.0035	\\
                    &       &  0.25 &   0.700873   &  0.1213    & 0.0047	\\
                    &       &  0.30 &   0.674033   &  0.1481    & 0.0058	\\
                    &       &  0.35 &   0.646090   &  0.1760    & 0.0066	\\
                    &       &  0.40 &   0.616896   &  0.2052    & 0.0067	\\
                    \hline
            \end{tabular}
            \label{tab:hehb2}	
    \end{table}

\begin{figure}
            \centering
            \includegraphics[width=.48\textwidth,angle=0]{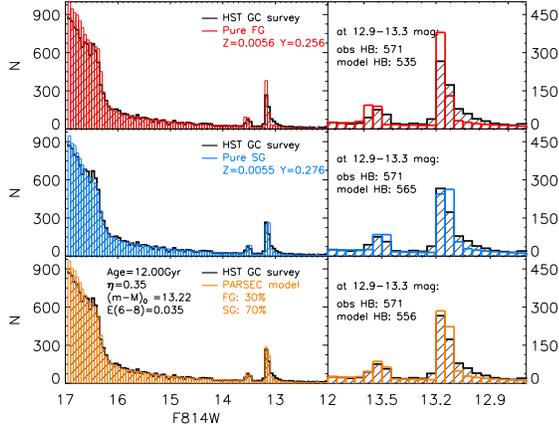}
            \caption{Comparison between the luminosity function of 47~Tuc data \citep{acsgc} 
            and that derived from the  new \parsec models,
            from the turn-off to the HB.
            The Y axis represent the star counts in  magnitude F814W.
            The black histogram filled with oblique lines is the data LF,
            whilst the red histogram in the upper panel,
            blue histogram in the middle panel,
            and orange histogram in the lower panel,
            represent 100\% FG of 47~Tuc [Z=0.0056, Y=0.256],
            100\% SG [Z=0.0055, Y=0.276],
            and their mix with 30\% from the FG and 70\% from the SG, respectively.
            The three panels on the right side show the LF of the RGBB and the HB region, 
            for each population mixture.
            The fitting parameters are: $\eta$=0.35,
            age=12 Gyr, (m$-$M)$_0$=13.22, and E(6-8)=0.035.}
            \label{fig:lf05_2}	
    \end{figure}

        The LFs from the turn-off to the HB with a RGB mass loss parameter
        $\eta=0.35$ are displayed in Fig.~\ref{fig:lf05_2} for  our best
        fitting parameters.  For comparison, the LF of HST GC survey data
        \citep{acsgc} is also plotted (black histogram filled with oblique
        lines) with the same binsize 0.05 mag.  We adopt the Salpeter IMF
        \citep{Salpeter1955} to generate LFs though, as discussed in
        Sec.~\ref{subsuc:eov}, LFs in this phase are not sensitive to IMF
        because the stellar mass varies  very little.  LFs are instead
        sensitive to the evolution time along the phase.  All model LFs are
        normalized to the total number of observed RGB stars within a range of
        F814W magnitude between 14~mag~-~16~mag.	The left panels of
        Fig.~\ref{fig:lf05_2} show the LFs from the turn-off to the HB, for a
        100\% FG (red histogram), a 100\%  SG (blue histogram), and the
        percentage adopted in Sec. \ref{subsuc:eov},  30\% from FG and 70\%
        from SG (orange histogram), respectively.  With our best isochrone
        fitting parameters, age=12.00~Gyr, (m$-$M)$_0$=13.22,  E(6-8)=0.035,
        $\eta$=0.35, and the  population percentage obtained from literature
        \citep{Carretta2009, Milone2012}, the model LF (orange histogram in
        each figure) show a very  good agreement with the observed LF.  The
        three right panels in  Fig.~\ref{fig:lf05_2}  are zoomed in on the HB
        and  RGBB regions.  The total number of HB stars within 12.9~-~13.3~mag
        in the observations and in the models are listed in the figure.  Since
        the LF is directly proportional to the evolution time, the good
        agreement of LF between model and observation in Fig.~\ref{fig:lf05_2}
        indicates that the hydrogen shell burning lifetime is correctly
        predicted in \parsec.

     \section{Comparison with other models and GC data}
     \label{sec:com}

   The new \parsec~$\alpha$~enhanced isochrones provide a very good fit of the
   color magnitude diagram of 47~Tuc in all evolutionary stages from the lower
   main sequence to the HB.  The location of the RGB bump shows that the
   efficiency of the envelope overshoot is quite significant, requiring EOV of
   $\Lambda_e=0.5H_P$. This can be considered a calibration of this phenomenon.
   We now use the calibrated EOV value to obtain $\alpha$-enhanced isochrones
   of different metallicities.  For this purpose we adopt the partition of
   heavy elements of the two stellar generations of 47~Tuc (\alpfe $\sim$ 0.4
   and 0.2).  In this section we compare our new $\alpha$-enhanced models with
   isochrones from other stellar evolution groups  and GC data of different
   metallicities.

\subsection{Comparison with other models}
     \label{subsec:other}

   The RGBB of globular clusters, as already said in section \ref{subsuc:eov},
   has been studied over 30 years since the 47~Tuc RGBB was observed in 1985
   \citep{king1985}.  However, there is a discrepancy between the observed
   brightness of RGBB and the model predictions: the model RGBB magnitude is
   about 0.2\textendash0.4 mag brighter than the observed one \citep{Fusi90,
   Cecco2010,Troisi2011}.  This discrepancy becomes more pronounced in
   metal-poor GCs \citep{Cassisi2011}.

\begin{figure*}
             \centering
             \includegraphics[width=.85\textwidth,angle=0]{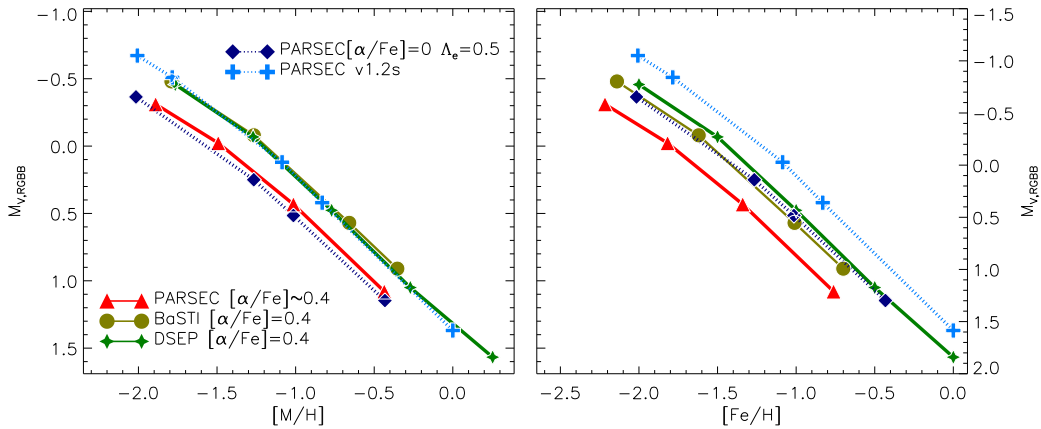}
             \caption{Comparison of the RGBB magnitude of different evolutionary tracks at 13Gyr
             as a function of [M/H] (left panel) and [Fe/H] (right panel).
             There are three different $\alpha$-enhanced models (\alpfe=0.4) in the figure:
             \parsec (red solid line with triangle), BaSTI (yelow green solid line with dot), 
             and DSEP model (green solid line with star).
             Other two sets of solar-scaled \parsec models (\alpfe=0) are  plotted for comparison:
             \parsec \texttt{v1.2S} with negligible overshoot (light blue dotted line with cross)
             and \parsec with EOV calibration $\Lambda_e=0.5H_p$ (dark blue dotted line with diamond).
             The Y axis is the mean value of the absolute F606W magnitude  of the RGBB (M$_{V,RGBB}$).}
             \label{fig:commod}
     \end{figure*}

    Here we compare the RGBB magnitude of our newly calibrated \parsec models
    with other $\alpha$-enhanced stellar tracks.  Since the BaSTI
    \citep{Pietrinferni2006, Pietrinferni2013} and DSEP \citep{Dotter2007a,
    Dotter2008} isochrones are publicly available online, we download the
    \alpfe=0.4 isochrones at 13 Gyr from BaSTI Canonical Models database and
    DSEP web tool 2012 version.  We then compare the mean values of their
    absolute RGBB magnitude in the F606W (HST ACS/WFC) band, with our models.
    Fig.  \ref{fig:commod} shows this comparison as a function of total
    metallicity [M/H] and iron abundance [Fe/H].  The model [M/H]  is
    approximated by:
     \begin{equation}
             [M/H] \approx \log \frac{Z/X}{Z_\odot/X_\odot}
             \label{eq:mh}
     \end{equation}
     And for both of the two new  \parsec models with $\alpha$ enhancement,
     [Fe/H] $\approx$ [M/H]-0.33.  The solar metallicity in \parsec is
     $Z_\odot=0.01524$ and  $Z_\odot/X_\odot=0.0207$.  Since DSEP models do not
     provide [M/H] directly  but only [Fe/H] in their isochrones, we calculate
     [M/H] following Eq.\ref{eq:mh} with total metallicity Z, He content Y, and
     solar $Z_\odot/X_\odot$  taken from their models.  Additionally, two
     \parsec models with solar-scaled metal mixture (\alpfe=0), \parsec
     \texttt{v1.2S} and  \parsec with EOV calibration from this work
     $\Lambda_e=0.5H_p$, are also plotted.  Compared with the new set of
     solar-scaled \parsec model with  $\Lambda_e=0.5H_p$ (dark blue line with
     diamond), the $\alpha$-enhanced one (red line with triangle) at the same
     [M/H] (thus same Z and Y) is  slightly brighter as we have already
     discussed in Sec. \ref{sec:input}.  We notice that the RGBB behavior of
     \parsec \texttt{v1.2S} in this figure is different from  Figure 3 of
     \citet{Joyce2015}, which compares \parsec \texttt{v1.2S} with other
     models.  The reason for this disagreement is unclear to us.

   Among the factors that may affect the brightness of the RGBB, as summarized
   in Sec. \ref{subsuc:eov}, we list the mixing efficiency and He contents.
   The helium-to-metal enrichment law of the different models are different, as
   discussed in Sec. \ref{sec:input}.  \parsec ($Y=0.2485 + 1.78Z$)  uses a
   slightly higher He abundance ($\sim$0.002) than the other two models (BaSTI:
   $Y=0.245+1.4Z$, DSEP: $Y=0.245+1.54Z$).  Different model also adopts
   different mixing length parameters.  The \parsec mixing length parameter is
   $\alpha_{MLT}=1.74$, BaSTI uses  $\alpha_{MLT}=1.913$, and DSEP adopts
   $\alpha_{MLT}=1.938$.  If all other parameters are the same, a higher He
   content and a smaller mixing length parameter lead to a brighter RGBB
   \citep{fu2016}.  This can explain why in the left panel of
   Fig.~\ref{fig:commod} the solar scaled \parsec \texttt{v1.2S} shows nearly
   the same M$_{V,RGBB}$ independent of [M/H] as BaSTI and DSEP models.
   \parsec \texttt{v1.2S} has slightly higher He content and smaller mixing
   length parameter which make the RGBB brighter as already discussed above,
   while its solar scaled metal mixture leads to a fainter RGBB at the same
   metallicity.  The combined effects  make the three models to show similar
   RGBB magnitude.  BaSTI and DSEP RGBB have almost the same performance and
   are eventually brighter than \parsec \alpfe$\sim$0.4 models no matter as a
   function of [M/H] or [Fe/H].  We remind that a fainter RGBB magnitude can be
   produced by a more efficient EOV, our new $\alpha$-enhanced models are
   computed with the calibrated EOV parameter  while,  BaSTI and DSEP do not
   consider envelope overshooting.  Also, compared to \parsec \texttt{v1.2S}
   ($\Lambda_e=0.05H_p$) the new solar scaled model with $\Lambda_e=0.5H_p$
   shifts M$_{V,RGBB}$ down by about 0.35 mag.  This brightness change is
   consistent with the work of \citet{Cassisi2002} who conclude that the
   difference should be of about 0.8 mag/$H_p$.  Since the RGBB brightness
   difference between the new \parsec $\alpha$-enhanced models and the solar
   scaled models with the same $\Lambda_e$ is much smaller than the difference
   between the two solar scaled \parsec models with different $\Lambda_e$, we
   conclude that the mixing efficiency has much stronger impact on the RGBB
   performance than the metal partition.

    \subsection{Comparison with other GC data}
     \label{subsec:gc}		

       Comparing the location of the RGB bump predicted by the models with the
       observed one in GCs with different metallicity is a good way to test the
       models.  

\begin{figure*}
             \centering
             \includegraphics[width=.85\textwidth,angle=0]{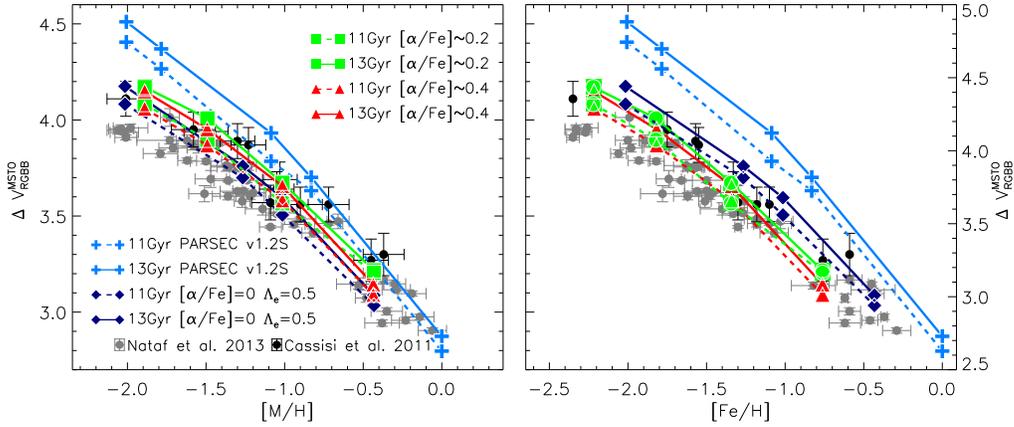}
             \caption{F606W magnitude difference between the  MSTO and the RGBB ($\Delta V^{MSTO}_{RGBB}$)
             as a function of the total metallicity [M/H] (left panel) and iron abundance [Fe/H].
             Four different sets of theoretical $\Delta V^{MSTO}_{RGBB}$ value are plotted, 
             at both 13Gyr (solid line) and 11Gyr (dashed line).
             Three of them are with new calibrated EOV $\Lambda_e=0.5H_p$:
             \alpfe $\sim$ 0.4 (red lines with triangle), \alpfe $\sim$ 0.2 (green lines with square),
             and \alpfe=0 (dark blue lines with diamond).
             Another one  is from the standard \parsec \texttt{v1.2S} (light blue lines with cross).
             The data are 55 clusters from \citet[][grey dots with error bar]{Nataf2013}
             and 12 clusters from \citet[][black dots with error bar]{Cassisi2011}. }
             \label{fig:msto}
     \end{figure*}

       In Fig.  \ref{fig:msto} we compare our new $\alpha$-enhanced models with
       HST data from \citet[][55 clusters]{Nataf2013} and  \citet[][12
       clusters]{Cassisi2011}.  The models extend till [M/H]$\sim -2$
       ([Fe/H]$\sim -2.3$).  For comparison, two sets of models with
       solar-scaled metal partition, \alpfe=0 (\parsec \texttt{v1.2S} and
       \parsec with $\Lambda_e=0.5H_p$) are also plotted.
     
     Here we use the magnitude difference between the RGBB and the main
     sequence turn-off (MSTO),  $\Delta V^{MSTO}_{RGBB}$, as a  reference  for
     comparison between the theoretical magnitude of RGBB and the observed one.
     Unlike the absolute magnitude $M_{V,RGBB}$, $\Delta V^{MSTO}_{RGBB}$ is
     not  affected by uncertainties in the distance modulus (m$-$M)$_0$ and
     extinction $A_V$ of the cluster.  There are also works using the magnitude
     difference between HB and RGBB \citep[$\Delta V^{RGBB}_{HB} = M_{V,RGBB} -
     M_{V,HB}$, e.g.][]{Fusi90, Cassisi1997,Cecco2010} or the one between HB
     and MSTO \citep[$\Delta V^{HB}_{TO} = M_{V,TO} - M_{V,HB}$,
     e.g.][]{VandenBerg2013} as a way to avoid distance and extinction
     uncertainties, but, as we have elaborated in Sec. \ref{subsec:hb}, the RGB
     mass loss together with different metal mixture and He content may affect
     the HB magnitude and thus make  $\Delta V^{RGBB}_{HB}$ difficult to be
     interpreted.  The only free parameter of the $\Delta V^{MSTO}_{RGBB}$
     method is the  age, if the composition of the cluster is fixed.  In
     Fig.~\ref{fig:msto} we compare the theoretical $\Delta V^{MSTO}_{RGBB}$
     value at typical GC ages of  11~Gyr and 13~Gyr, with the observed value
     from \citet{Nataf2013} and \citep{Cassisi2011}.  The comparisons are
     displayed both in the [M/H] frame and [Fe/H] frame.  The MSTO in
     \citet{Nataf2013} is defined by taking the bluest point of a polynomial
     fit to the upper main sequence of each GC in the (F606W, F606W$-$F814W).
     \citep{Cassisi2011} derive the MSTO magnitude by fitting isochrones to the
     main sequence.  To obtain  the theoretical MSTO F606W magnitude in our
     model,  we select the bluest point of the isochrone in the main sequence.
     The models of 13~Gyr show larger difference between RGBB and MSTO $\Delta
     V^{MSTO}_{RGBB}$ than those at 11~Gyr.  Models with \alpfe$\sim$0.2 show a
     slightly greater $\Delta V^{MSTO}_{RGBB}$ value than the models computed
     with \alpfe$\sim$0.4.  In the right panel of the figure we see that  the
     $\alpha$-enhanced models show greater  $\Delta V^{MSTO}_{RGBB}$ than the
     solar scaled ones with the same [Fe/H].  This said, if one has [Fe/H]
     measurement of a GC with $\alpha$ enhancement and takes  $\Delta
     V^{MSTO}_{RGBB}$ as an age indicator, choosing the solar scaled models
     will lead to an underestimated  cluster age.

    Compared to the previous \parsec version \texttt{v1.2S},  the new models
    significantly improve the $\Delta V^{MSTO}_{RGBB}$ prediction in both the
    [M/H] frame and [Fe/H] frame.  At the most metal-poor side, around
    [M/H]$\sim-2.0$ ([Fe/H]$\sim -2.3$), the new models are consistent with
    \citet{Cassisi2011} data (black dots), but are  higher than the values
    derived by \citet{Nataf2013} (grey dots) by $\sim$0.1 mag.  We will discuss
    the possible reasons in the next section.

\section{Summary and discussion}
\label{sec:conc}

Studies on globular clusters, Galactic bulge, halo, and thick disk call
for stellar models with $\alpha$ enhancement because stars residing in
them have $\alpha$-to-iron number ratio larger than the solar value.
This ratio, \alpfe, not only affects the stellar features like the
luminosities and effective temperature, but also echoes the formation
history of the cluster/structure the stars are in.  To investigate such
stars, and to trace back their formation history, we have now extend the
\parsec models to include $\alpha$-enhanced mixtures. 

In this paper we  check the $\alpha$-enhanced models with the nearby globular
cluster 47~Tuc (NGC 104).  The chemical compositions including the helium
abundances of 47~Tuc are studied  by many works.  We collect detailed elemental
abundances of this cluster and derive absolute metal mixtures for two
populations: first generation [Z=0.0056, Y=0.256], and second generation
[Z=0.0055, Y=0.276].  The  $\alpha$-to-iron ratio of them are \alpfe=0.4057
(\alpfe$\sim$0.4) and \alpfe=0.2277 (\alpfe$\sim$0.2), respectively.  We
calculate evolutionary tracks and isochrones with these two  $\alpha$-enhanced
metal mixtures, and fit color-magnitude diagram to HST/ACS data.  The model
envelope overshooting is then calibrated to the value  $\Lambda_e=0.5 H_P$ in
order to reproduce the RGB bump morphology in 47~Tuc.  After the calibration,
the new $\alpha$-enhanced isochrones nicely fit the data from the low main
sequence to the turn-off, giant branch, and the horizontal branch with age of
12.00~Gyr, distance modulus (m$-$M)$_0$=13.22 ( (m$-$M)$_{F606W}$=13.32) , and
reddening E(6-8)=0.035.  These results compare favorably with many other
determinations in the literature.  The luminosity functions inform us that the
lifetime of hydrogen  burning shell appears to be  correctly predicted.  By
studying the  morphology of the horizontal branch, we conclude that the mean
mass lost by stars during the RGB phase is around  0.17 \msun .

\begin{figure}
         \centering
         \includegraphics[width=.48\textwidth,angle=0]{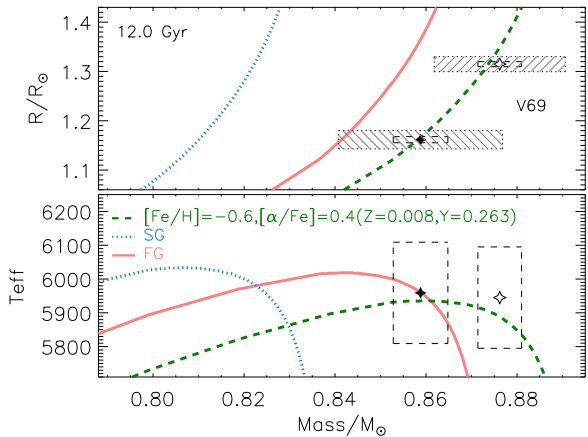}
         \caption{ Mass-radius and Mass-Teff diagrams for binary V69.
         The observed values of the two components of V69 are marked with black stars,
         their corresponding 1$\sigma$ and 3$\sigma$ uncertanties are indicated
         with dashed line boxes and shaded boxes, respectively \citep[data from][]{Thompson2010}.
         Isochrones of 47 Tuc FG (red solid lines) and SG (blue dotted lines) are overlaid.
         To illustrate the effects of a higher metallicity,
         isochrones of [Fe/H]=-0.6, [$\alpha$/Fe]=0.4 (Z=0.008, Y=0.263) 
         are also displayed (green dashed lines).
         All isochrones are with our best estimate age 12.0 Gyr. 
         }
         \label{fig:v69}
 \end{figure}

There are also other methods to estimate the  age of this cluster in the
literature.  For instance, mass-radius constraints of the detached eclipsing
binary stars V69 in 47 Tuc have also been used \citep{Weldrake2004,
dotter2008p, Thompson2010, Brogaard2017}.  This approach, which considers the
two components of the detached binary as single stars, is much less affected by
the uncertainties arising from the unknown distance, reddening and
transformation from the theoretical to the observational plane.
\citet{Thompson2010} derive an age of
11.25$\pm$0.21(random)$\pm$0.85(systematic) for 47 Tuc, and
\citet{Brogaard2017} give 11.8 Gyr as their best estimate with 3$\sigma$ limits
from 10.4 Gyr to 13.4 Gyr.  We  examined the mass-radius and mass-Teff
constraints provided by V69 with our best fit FG and SG  models at 12.0 Gyr, as
shown in Fig. ~\ref{fig:v69}.  We have adopted for the two components of V69
the following data \citep{Thompson2010}: current mass $M_p$= 0.8762$\pm$0.0048
~\msun, $M_s$=0.8588$\pm$0.0060 ~\msun, radius $R_p$=1.3148$\pm$0.0051~
$R_\odot$, $R_s$=1.1616$\pm$0.0062~ $R_\odot$ and effective temperature
Teff$_p$=5945 $\pm$150~K,  Teff$_s$=5959 $\pm$150~K of the primary and
secondary star, respectively.  Differences between the FG and SG isochrones are
due to the difference of He abundances and [$\alpha$/Fe].  The comparison with
our models indicate that V69 cannot belong to the SG population.  Concerning
the mass-radius relation, which provides the most stringent constraints on the
two stars, we see that our best fit FG isochrone of 12 Gyr is only marginally
able to reproduce the secondary component (filled black star), within
3$\sigma$, while there is a tension with the radius of the primary component
(empty star).  A better match can be obtained for both components by
considering a slightly higher metallicity, as also suggested by
\citet{Brogaard2017}.  For example, the green dashed isochrones in Fig.
~\ref{fig:v69} illustrate the effects of assuming  [Fe/H]=-0.6 and
[$\alpha$/Fe]=0.4, that correspond to Z=0.008,  and an He content of Y=0.263.
In this example the He content of the isochrone follows Equ.~\ref{eq:he} and
the [$\alpha$/Fe] value is chosen to be 0.4 without any special consideration.
We also note that, to bring the radius discrepancy of the primary component
within 3$\sigma$ or 1$\sigma$, with the abundances assumed for the FG
population, an age of 11.7~Gyr or 11.2~Gyr would be required,  respectively.
These ages, in particularly the lower one, would be quite different from the
one obtained by the best fit presented in the paper.  Up to now we have assumed
that effects of binary interaction are negligible so that V69 can be analysed
with single star evolution models.  However considering other possible causes
for the discrepancy, we note that the primary component of V69 is among the
bluest stars past the turn-off of the CMD \citep[see Fig. 4 of][]{Thompson2010}
so that the fitting isochrone would likely fit the bluer envelope of the
cluster CMD.  We thus suspect that its position in the CMD could partly be due
to the effects of the binary dynamical interaction with the companion star.  In
fact, tidal effects in close binary stars may change the structure and
evolution of stars even before any possible mass transfer \citep{demink2009,
Song2016}.  In particular, the primary star of V69 matches our model that shows
a thin surface convective envelope ($\sim$2\% of of total mass) and, following
the simple approximation of  \citet[][equation 6.1]{Zahn1977}, its tidal
synchronization time should be roughly $\sim$7.2~Gyr, comparable to its current
age.  Tidal friction during the previous evolution may have introduced shear
mixing and extra turbulence \citep{Lanza2016} at the base of the external
convective region, whose effects are primarily those of mitigating Helium and
heavy element diffusion away from the convective region.  The net effect will
be a lower growth of the surface hydrogen abundance with a corresponding
decrease of the current surface opacity, that should result in a smaller
current radius, in the observed direction.  Of course a more sophisticated
theoretical analysis is necessary to assess if the location of the primary star
of V69 in the CMD might be affected by previous dynamical interaction, but this
is beyond the scope of this paper.

The envelope overshooting calibration together with the $\alpha$-enhanced metal mixtures of 47~Tuc are applied
to other metallicities till Z=0.0001.
The RGB bump magnitudes of the new  $\alpha$-enhanced isochrones are compared with other stellar models and globular cluster observations.
We take $\Delta^{MSTO}_{RGBB}$, the magnitude difference between the main sequence turn-off and the RGB bump,
as the reference to compare with the observation in order to avoid uncertainties from distance  and extinction.
Our new models fit the data quite well and significantly improve the prediction of RGB bump magnitude compared to previous models.

      \begin{figure}
              \centering
              \includegraphics[width=.48\textwidth,angle=0]{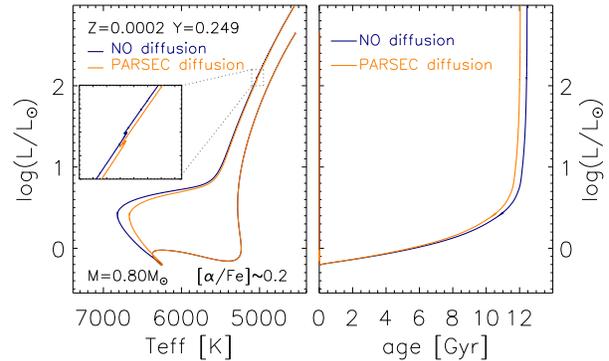}
              \caption{Comparison of [Z=0.0002 Y=0.249] tracks 
              for a M=0.80 \msun~ star with (orange line) and without (blue line) diffusion.
              The left panel shows the HRD of these two tracks, with the RGBB region zoomed in in the sub-figure.
              The right panel illustrates the luminosity evolution as the star ages.}
              \label{fig:dif}
      \end{figure}

      However we notice that in Fig.~\ref{fig:msto} around [M/H]=$-2.0$
      ([Fe/H]$\sim -2.3$) our model predicts  $\Delta V^{MSTO}_{RGBB}$ about
      $\sim$0.1 mag greater than the data points obtained by \citet{Nataf2013}.
      If we consider a more He-rich model with the same metallicity Z, the
      discrepancy will become even larger.  There are works arguing  that
      diffusion also affects the brightness of RGBB
      \citep[eg.][]{Michaud2010,Cassisi2011, Joyce2015}.  \citet{Michaud2010}
      conclude that without atomic diffusion the RGBB luminosity is about $0.02
      ~dex$ brighter ($\sim$ 0.05 mag).  In~\parsec we always take diffusion
      into account.  To see the effect of diffusion on RGBB morphology, we
      calculate a  0.80 \msun~ model without diffusion and compare it with the
      one with standard \parsec diffusion in Fig.~\ref{fig:dif}. The two stars
      have the same metallicity, He content, stellar mass, and \alpfe.  As we
      can see from the right panel   diffusion shorten the main sequence
      life-time.  The left panel of this figure is HRD, similar to that in
      Figure~1 of  \citet{Michaud2010}, evolutionary track with diffusion shows
      redder MSTO and slightly fainter RGBB.  For isochrones obtained from
      these two sets of evolutionary tracks, at 13 Gyr the RGBB without
      diffusion is 0.072 mag (in F606W) brighter than the one with diffusion,
      and $\Delta V^{MSTO}_{RGBB}$ value is 0.008 mag (in F606W) larger.  This
      result confirms that inhibiting the diffusion during H-burning phase will
      eventually makes the discrepancy more severe.  \citet{Pietrinferni2010}
      conclude that the updated nuclear reaction rate for
      \reac{N}{14}{p}{\gamma}{O}{15} makes RGBB brighter by $\sim$0.06 mag
      compared to the old rate.  However we remind that we are already adopting
      the new rate \citep{im05} for this reaction (Table \ref{tab:reac}).
      Assuming that all other input physics, in particular opacities, is
      correct, the only possible solution to cover this $\sim$0.1 mag
      discrepancy is that the mixing at the bottom of the convective envelope
      is even higher than that assumed here.  Either EOV in metal-poor stars is
      larger than our adopted value \citep[e.g $\Lambda_e = 0.7H_p$ suggested
      by][]{Alongi1991} or  another kind of extra mixing is responsible. 

     \begin{figure*}
             \centering
             \includegraphics[width=.85\textwidth,angle=0]{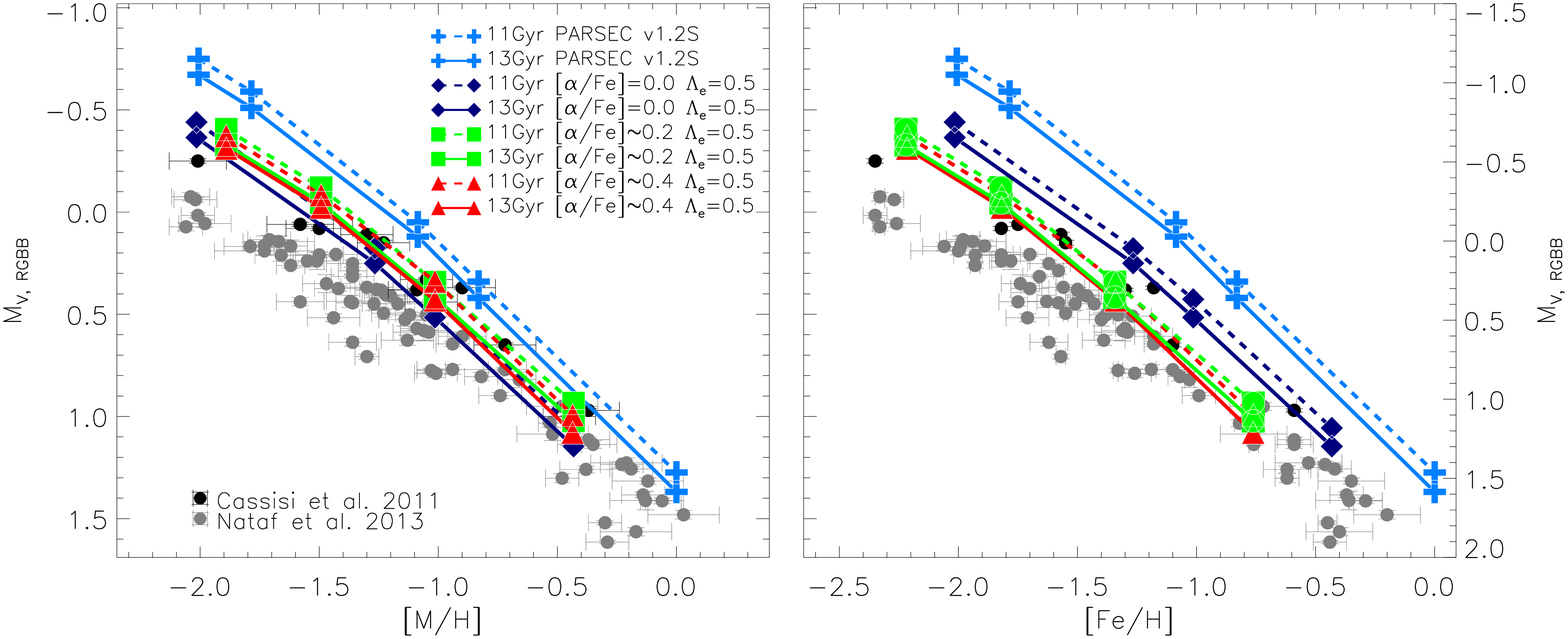}
             \caption{Absolute  magnitude in F606W band of RGBB (M$_{V,RGBB}$)
             as a function of metallicity [M/H] (left panel) and iron abundance [Fe/H] (right panel).
             Three sets of EOV-calibrated \parsec ~models, \alpfe=[0.0, 0.2, 0.4],
             and \parsec \texttt{v1.2S} are shown both at
             age 13Gyr (solid line) and 11Gyr (dashed line).
             For comparison, \citet[][grey filled dots with error bar]{Nataf2013} 
             and \citet[][black filled dots with error bar]{Cassisi2011} RGBB data 
             are plotted. 
             See the text for the details.}
             \label{fig:combump}
     \end{figure*}

    We have shown in Sec. \ref{subsec:gc}	that,
    comparing the absolute magnitude $M_{V,RGBB}$ between model and observation directly,
    would introduce uncertainties from distance and extinction.
    However putting  the $M_{V,RGBB}$  and  $\Delta V^{MSTO}_{RGBB}$ comparison together, 
    could help us to constrain the distance  of the clusters.
    Fig. ~\ref{fig:combump} displays the differences between the theoretical $M_{V,RGBB}$ 
    and those of data from \citet{Nataf2013} and \citet{Cassisi2011}. 
    The absolute magnitude   $M_{V,RGBB}$ of the data takes into account the apparent distance modulus, 
    hence extinction effects are excluded.
    The four sets of models  in Fig.~\ref{fig:combump} are the same as those in Fig.~\ref{fig:msto}.
    The  discrepancy of $M_{V,RGBB}$ between models and the data at the metal-poor end in Fig.~\ref{fig:combump} is larger
    than that of $\Delta V^{MSTO}_{RGBB}$ in Fig.~\ref{fig:msto}
    in both [M/H] and [Fe/H] frames.
   Since $\Delta V^{MSTO}_{RGBB}$ is a reference without distance effect,
   this larger discrepancy  indicates that the apparent distance modulus (m$-$M)$_V$ used in Fig.~\ref{fig:combump} \citep{Nataf2013}
   for metal-poor GCs   are underestimated.

      In a following paper of the ``\parsec $\alpha$-enhanced stellar evolutionary tracks and isochrones''   series,
      we will provide other \alpfe choices
      based on metal mixtures derived from ATLAS9 APOGEE atmosphere models \citep{Meszaros2012}.
      The full set of isochrones with chemical compositions suitable for GCs and Galactic bulge/thick disk stars
      will be available online after the full calculation and calibration are performed.

\section{Acknowledgements}

XF thanks Angela Bragaglia and Francesca Primas for the useful discussion,
and Fiorella Castelli for the guide on ATLAS12 calculation.
AB acknowledges PRIN INAF 2014 'Star formation and evolution in
galactic nuclei',
and thanks Michela Mapelli for the useful discussion on binary interaction.
PM, JM, YC and AN acknowledge support from the ERC Consolidator Grant funding scheme (project STARKEY,
G. A. n. 615604).

\bibliographystyle{mn2e/mn2e_new} 
\bibliography{47tuc}
 \label{lastpage}

\clearpage

\end{document}